\documentclass[a4paper,11pt]{article}
\usepackage[margin=0.5in]{geometry} % full-width
\usepackage{enumerate}
\usepackage{jcappub}
\usepackage{aas_macros}
\usepackage{amsmath}
\usepackage{amssymb}
\usepackage{bm}
\usepackage{natbib}
\usepackage{xspace}
\usepackage{lipsum}
\usepackage[utf8]{inputenc}
\usepackage[T1]{fontenc}
\usepackage{hyperref}
\usepackage{caption}
\usepackage{verbatim}
\usepackage[dvipsnames]{xcolor}

\hypersetup{
	unicode,
	pdfauthor={Author One, Author Two, Author Three},
	pdftitle={Catalog-based pseudo-$C_\ell$s},
	pdfsubject={Catalog-based pseudo-$C_\ell$s},
	pdfproducer={LaTeX},
	pdfcreator={pdflatex}
}
\usepackage{graphicx, color}
 
\newcommand{\planck}{{\sl Planck}\xspace}
\newcommand{\quaia}{\textit{Quaia}\xspace}
\newcommand{\gaia}{{\sl Gaia}\xspace}

\newcommand{\nv}{\hat{\bf n}}
\newcommand{\dint}{\,\text{d}}
\newcommand{\wtj}[6]{\left(\begin{array}{ccc} #1 & #2 & #3\\#4 & #5 & #6\end{array} \right)}
\newcommand{\nmt}{{\tt NaMaster}\xspace}
\newcommand{\hp}{{\tt HEALPix}\xspace}

\title{\boldmath Catalog-based pseudo-$C_\ell$s}

\author[a]{Kevin Wolz,}
\author[a]{David Alonso,}
\author[b]{Andrina Nicola}

\affiliation[a]{Department of Physics, University of Oxford, Denys Wilkinson Building, Keble Road, Oxford OX1 3RH, United Kingdom}
\affiliation[b]{Argelander Institut f\"ur Astronomie, Universit\"at Bonn, Auf dem H\"ugel 71, 53121 Bonn, Germany}

\emailAdd{kevin.wolz@physics.ox.ac.uk}

\abstract{We present a formalism to extract the angular power spectrum of fields sampled at a finite number of points with arbitrary positions -- a common situation for several catalog-based astrophysical probes -- through a simple extension of the standard pseudo-$C_\ell$ algorithm. A key complication in this case is the need to handle the shot noise component of the associated discrete angular mask which, for sparse catalogs, can lead to strong coupling between very different angular scales. We show that this problem can be solved easily by estimating this contribution analytically and subtracting it. The resulting estimator is immune to small-scale pixelization effects and aliasing, and, more interestingly, unbiased against the contribution from measurement noise uncorrelated between different sources. We demonstrate the validity of the method in the context of cosmic shear datasets, and showcase its usage in the case of other spin-$0$ and spin-$1$ astrophysical fields of interest. We incorporate the method in the public \href{https://github.com/LSSTDESC/NaMaster}{\nmt} code.}
\begin{document}
\maketitle
\flushbottom

\section{Introduction}
  In cosmology and astrophysics, a plethora of observable quantities can be described by projected fields on the sky. These fields may be observed and mapped in a continuous manner, as is the case of Cosmic Microwave Background (CMB) observations, or may be observed only at the discrete positions of a finite number of sources. Depending on the physical nature of the quantity at hand, these quantities will transform differently under rotations of our angular coordinates, leading to spin-0 fields (i.e., scalars, such as the CMB temperature anisotropies, the lensing convergence, or the galaxy overdensity), spin-1 fields (i.e., vectors such as transverse peculiar motions or lensing displacements), spin-2 fields (i.e., tensors such as the CMB polarization anisotropies or weak lensing shear), and higher-spin fields (e.g. the spin-4 polarization of gravitational waves). The angular power spectrum, corresponding to the covariance between angular fluctuations in two different fields at a given multipole scale $\ell$, is one of the most used summary statistics in cosmology and astrophysics and, in particular, encodes all available information in the case of Gaussian fields. For this reason, being able to accurately estimate the angular power spectrum is a key requirement for many cosmological probes. In the case of galaxy weak lensing, for instance, a precise and accurate reconstruction of the cosmic shear power spectrum on sub-arcminute scales is of vital importance to obtain meaningful constraints on the gravity-dominated, nonlinear regime of cosmic structure formation, scrutinizing the Universe’s matter content and nature of Dark Energy \citep{astro-ph/0609591, 1201.2434}. This is a key goal for current and next-generation galaxy imaging surveys, such as the Kilo Degree Survey (KiDS)\footnote{\url{http://kids.strw.leidenuniv.nl}}, the Dark Energy Survey (DES)\footnote{\url{https://www.darkenergysurvey.org}}, the Hyper Suprime-Cam\footnote{\url{https://www.naoj.org/Projects/HSC/}}, Euclid\footnote{\url{https://www.euclid-ec.org}}, and the Rubin Observatory Legacy Survey of Space and Time (LSST)\footnote{\url{https://www.lsst.org}}. Valuable complementary insight may be gained from the angular power spectrum of other discretely sampled tracers of the matter density, such as the overdensity of galaxies \cite{1912.08209,2111.09898,2312.12285}, fluctuations in their sizes \cite{astro-ph/0606672,2209.11063}, fluctuations in the luminosity distance of standard sirens \cite{2208.05959}, transverse peculiar motions \cite{2305.15893}, or the dispersion measure of extragalactic Fast Radio Bursts (FRBs) \citep{1901.02418,2103.14016,2201.04142,2302.10072}.

  There is a large literature of unbiased power spectrum estimators that are optimal, have well-understood statistical properties, and robustly meet common observational challenges such as partial sky coverage, linear sky contamination, and inhomogeneous noise \citep{astro-ph/9611174, astro-ph/9708203, astro-ph/9712121, astro-ph/0307515, astro-ph/0503603, astro-ph/0503604, astro-ph/0012120}. For small angular scales, optimal quadratic estimators are usually computationally prohibitive, and the most common alternative is their simpler incarnation, the so-called ``pseudo-$C_\ell$'' estimator \citep{astro-ph/0105302, astro-ph/0410394, 1809.09603}. The pseudo-$C_\ell$ estimator effectively assumes a diagonal covariance matrix for the data, and is therefore close to optimal for spatially uncorrelated or white-noise-dominated data. As a consequence of this assumption, the mode-coupling induced by the spatial variation of the resulting inverse-variance weights (i.e., the ``mask'') can be estimated using fast, analytical methods. Variations of the method allow it to account for the impact of linear contaminant deprojection, and for the effects of $E$- and $B$-mode purification in non-scalar fields \cite{1609.03577,1809.09603}.

  In its standard implementation, the pseudo-$C_\ell$ method takes, as input, pixelized maps of the fields in question. When the original observations are in the form of discrete measurements at arbitrary sky positions, as is the case for any catalog-based observable, this results in two fundamental problems. First, generating a pixelized map from the discrete sources induces a bias at scales comparable to, and smaller than, the pixel size. The form of this bias depends on a combination of pixel smoothing and aliasing of small-scale power, and is non-trivial to characterize in general \cite{2010.09717}. Second, when the source number density is low, the mask is dominated by Poisson noise, also known as shot noise, which, as we will see, leads to statistical couplings between distant angular multipoles. If not treated carefully, this can lead to strong numerical instabilities and catastrophic biases in the estimator. The issue of small-scale power in the mask has been traditionally handled via apodization and smoothing of the mask edges \cite{astro-ph/0105302,astro-ph/0303414,1809.09603}, but this approach is highly impractical in the context of catalog-based observations for sparse samples.

  This work addresses the issue of how to define an unbiased power spectrum estimator for fields of arbitrary spin discretely sampled on the sphere. Building on the recent work on this topic by \cite{2312.12285}, we forgo the pixelization step completely and directly estimate spherical harmonic transforms of the field and its measurement weights at the discrete source positions. This allows us to build a pixel-independent version of the pseudo-$C_\ell$ estimator, addressing the first problem described above. We address the second problem (shot noise instabilities) by introducing a simple modification to the estimator that accounts for uncorrelated noise, both in the mask and the field, and simultaneously eliminates both the associated noise bias and the numerical instabilities of the mask-induced mode coupling. More specifically, the contribution to mode coupling from the mask shot noise is estimated analytically and subtracted from the measured pseudo-$C_\ell$, together with a na\"ive, data-based estimate of the noise bias. As we will show, the resulting estimator is then immune to any bias from uncorrelated noise, and benefits from a numerically stable mode-coupling matrix, independently of the number density of sources in the catalog. We implement this methodology in the publicly available \nmt package\footnote{\url{https://github.com/LSSTDESC/NaMaster/}}. As we were preparing this manuscript, we learned of related work presented in Tessore et al. 2024 \cite{TessoreInPrep}. The results there agree with those presented here.

  This paper is structured as follows. In Section \ref{sec:meth}, we lay out the formalism that defines our unbiased estimator. In Section \ref{sec:val}, we validate it on simulated cosmic shear catalogs with a realistic source distribution, based on real data from the Kilo Degree Survey, also demonstrating its validity on significantly lower number densities. In Section \ref{sec:ex}, we present several example applications of the methodology, including state-of-the-art cosmic shear data from the Dark Energy Survey, transverse proper motion surveys, and FRB-based dispersion measure maps. We conclude in Section \ref{sec:con}.

\section{Methods}\label{sec:meth}
  This section presents the formalism to estimate the angular power spectrum from sparse observations of a given field at a discrete set of points on the sphere using the pseudo-$C_\ell$ formalism. Section \ref{ssec:meth.pcl} reviews the main concepts behind the standard pseudo-$C_\ell$ method, and the application to discrete tracers is described in Section \ref{ssec:meth.disc}. We first present the formalism in the simpler case of scalar fields, and then, in Section \ref{ssec:meth.spin} present the generalization to spin-$s$ quantities.

  \subsection{Pseudo-$C_\ell$s}\label{ssec:meth.pcl}
    Let $f(\nv)$ be a scalar field on the sphere, and let $w(\nv)$ be a sky mask (often also called weights maps, or sky window function). The mask may be as simple as a binary map (0 for unobserved regions, 1 for observed regions) but, more usefully, it should represent the precision with which $f$ has been measured at different points (e.g. an inverse-variance map will be optimal for noise-dominated observations). Because of this, $w(\nv)$ is often referred to as a ``window function'' or ``weights map'', but we will stick to the simpler ``mask'' in what follows.

    The field is related to its harmonic coefficients $f_{\ell m}$ through a spherical harmonic transform (and its inverse):
    \begin{equation}
      f(\nv)=\sum_{\ell m} Y_{\ell m}(\nv)\,f_{\ell m}\hspace{6pt}\leftrightarrow\hspace{6pt}f_{\ell m}=\int d\nv\,f(\nv)\,Y^*_{\ell m}(\nv),
    \end{equation}
    where $Y_{\ell m}(\nv)$ are the spherical harmonic functions. The spherical harmonic coefficients of the mask, or any other sky map, are defined in the same way.

    Let $f^w(\nv)\equiv f(\nv)w(\nv)$ be the masked field. The ``pseudo-$C_\ell$'' of $f$ is simply the na\"ive estimate of the masked field's power spectrum ignoring the fact that it's masked:
    \begin{equation}
      \tilde{C}^f_\ell\equiv\frac{1}{2\ell+1}\sum_{m={-\ell}}^\ell f^w_{\ell m}f^{w*}_{\ell m}.
    \end{equation}
    We commonly assume the field to receive signal and noise contributions: $f=s+n$, where the signal $s$ is a statistically isotropic field, such that
    \begin{equation}
      \langle s_{\ell m}s^*_{\ell' m'}\rangle=\delta_{\ell\ell'}\delta_{mm'}S_\ell,
    \end{equation}
    here $S_\ell$ is the signal power spectrum, and is the quantity we are ultimately after.

    The key result of the pseudo-$C_\ell$ estimator is the relation between the ensemble average of $\tilde{C}^f_\ell$ and $S_\ell$:
    \begin{equation}\label{eq:pcl}
      \langle\tilde{C}^f_\ell\rangle=\sum_{\ell'}M^w_{\ell\ell'}S_{\ell'}+\tilde{N}_\ell,
    \end{equation}
    where $M^w_{\ell\ell'}$ is the mode-coupling matrix:
    \begin{equation} \label{eq:mcm_mask}
      M^w_{\ell\ell'}=\frac{2\ell'+1}{4\pi}\sum_{\ell''}\tilde{C}^w_{\ell''}\,(2\ell''+1)\wtj{\ell}{\ell'}{\ell''}{0}{0}{0}^2,
    \end{equation}
    here $\tilde{C}^w_\ell\equiv\sum_m |w_{\ell m}|^2/(2\ell+1)$ is the pseudo-$C_\ell$ of the mask. $\tilde{N}_\ell$ is the ``noise bias'', given by the ensemble average of the pseudo-$C_\ell$ of the masked noise:
    \begin{equation}\label{eq:pcl_noise}
      \tilde{N}_\ell\equiv\left\langle\frac{1}{2\ell+1}\sum_mn^w_{\ell m}n^{w*}_{\ell m}\right\rangle.
    \end{equation}
    Depending on the type of observations, and our understanding of the noise properties, $\tilde{N}_\ell$ may be estimated analytically, via Monte-Carlo simulations, or avoided by using cross-correlations between observations with the same signal but different noise realizations (e.g. CMB maps constructed from disjoint sets of timestreams). As we will see, in the case of discretely-sampled fields, the noise bias can be entirely avoided by construction under the assumption that the noise contributions from different points are uncorrelated. The expressions above may be generalized to the case of cross-correlations between different fields, by simply replacing $\tilde{C}^w_\ell$ with the cross-spectrum between the masks of both fields.

  \subsection{The case of discretely-sampled fields}\label{ssec:meth.disc}
    Consider now observations of the field $f$ only at a discrete set of $N$ points on the sphere. Most commonly, these will be the sky positions of sources in a catalog. Let $w_i$ be the weight associated to the $i$-th point, with sky position $\nv_i$, and $f_i\equiv f(\nv_i)$. To ease notation, in what follows, the subscript $_i$ applied to any function defined on the sphere will be shorthand for the value of that function at $\nv_i$. The mask and masked field can, in this case, be written as a sum over delta functions:
    \begin{equation}
      w(\nv)=\sum_i\,w_i\,\delta^D(\nv,\nv_i),\hspace{12pt}f^w(\nv)\equiv w(\nv)\,f(\nv)=\sum_i w_if_i\,\delta^D(\nv,\nv_i),
    \end{equation}
    where $\delta^D(\nv_1,\nv_2)$ is the Dirac delta function on the sphere\footnote{That is, $\delta^D(\nv_1,\nv_2)\equiv\delta(\varphi_1-\varphi_2)\delta(\cos\theta_1-\cos\theta_2)$, where $(\theta_n,\varphi_n)$ are the spherical coordinates of the unit vector $\nv_n$.}. The harmonic coefficients of $w$ and $f^w$ are then sums over the harmonic functions sampled at the positions of the discrete sources
    \begin{equation}\label{eq:sht_disc}
      w_{\ell m}=\sum_i w_i\,Y_{\ell m,i}^*,\hspace{12pt}f^w_{\ell m}=\sum_i w_if_i\,Y_{\ell m,i}^*.
    \end{equation}
    An efficient evaluation of such discrete harmonic transform over a non-regular set of points is, in general, computationally more expensive than standard approaches making use of regular pixelization schemes \cite{1303.4945}. Nevertheless, approaches for dealing with this problem have been recently developed \cite{2304.10431,2312.12285}, and we will use the implementation of such ``general'' spherical harmonic transforms built in the {\tt ducc} library\footnote{\url{https://mtr.pages.mpcdf.de/ducc}}, as described in \cite{2304.10431}. This paper will not discuss the details of the numerical implementation of these methods and instead will deal with the implementational details of the pseudo-$C_\ell$ algorithm itself, once these discrete transforms are available. We refer readers to \cite{2304.10431} further details.

    The pseudo-$C_\ell$ of both mask and masked field is given by:
    \begin{align}\label{eq:pcl_mask}
      &\tilde{C}^w_\ell=\sum_{ij}w_iw_j\,\frac{1}{2\ell+1}\sum_{m}Y_{\ell m,i}^*Y_{\ell m,j},\\
      &\tilde{C}^f_\ell=\sum_{ij}w_iw_jf_if_j\,\frac{1}{2\ell+1}\sum_{m}Y_{\ell m,i}^*Y_{\ell m,j}.
    \end{align}
    These equations may be further simplified making use of the addition theorem for spherical harmonics (Eq. \ref{eq:Ylm_add}). However, it will be more useful to leave them as is for now.

    \subsubsection{Noise bias from uncorrelated measurements}\label{sssec:meth.disc.noise}
      Let us now consider our measurements of the field $f=s+n$ to contain some common signal $s$ and noise component $n$, which we assume to be uncorrelated among different sources. Under this assumption, the noise contribution to $\tilde{C}^f_\ell$ reads
      \begin{equation}
        \langle n_i n_j\rangle = \delta_{ij}\sigma_{N,i}^2,
      \end{equation}
      where $\sigma_{N,i}^2$ is the noise variance in the $i$-th point. In this case, the noise bias only receives contributions from the variance at the same source (i.e., the zero-lag correlation function):
      \begin{equation}\label{eq:Nell_uncorr}
        \tilde{N}_\ell=\sum_{ij}w_iw_j\langle n_in_j\rangle\frac{1}{2\ell+1}\sum_mY^*_{\ell m,i}Y_{\ell m,j}=\frac{1}{4\pi}\sum_i w_i^2\sigma_{N,i}^2,
      \end{equation}
      where, in the last equality, we have made use of the addition theorem for spherical harmonics (Eq. \ref{eq:Ylm_add}) for the special case $i=j$. The noise bias is therefore ``white'' (i.e., scale-independent), and non-zero at all $\ell$s.

      Since the standard pseudo-$C_\ell$ estimator requires us to estimate $\tilde{N}_\ell$ and subtract it from $\tilde{C}^f_\ell$, in order to obtain an unbiased estimator of the signal power spectrum $S_\ell$, it is important for us to consider practical ways to estimate $\tilde{N}_\ell$. Assuming our measurements of $f$ to be dominated by noise at any single point, one potentially simple way to do this would be to estimate $\tilde{N}_\ell$ directly from the variance of the data itself. This methodology has been commonly applied, in the past, when estimating cosmic shear power spectra, where measurements are indeed dominated by noise \cite{2203.07128, PhysRevD.108.123519, 2403.13794}. %\AN{an example would be the DES SV cosmic shear analysis}
      In this situation, and looking at Eq. \ref{eq:Nell_uncorr}, we can thus propose the following estimator for $\tilde{N}_\ell$:
      \begin{equation}\label{eq:Nf}
        \tilde{N}^f\equiv\frac{1}{4\pi}\sum_i w_i^2f_i^2.
      \end{equation}
      Expanding the field measurements into signal and noise components, we can thus see that $\tilde{N}^f$ is a biased estimate of $\tilde{N}_\ell$ due to the presence of signal in our measurements:
      \begin{equation}
        \langle \tilde{N}^f\rangle=\tilde{N}_\ell+\frac{1}{4\pi}\sum_i w_i^2\langle s_i^2\rangle.
      \end{equation}
      For a statistically isotropic field, such as $s$, the field variance is related to its power spectrum via
      \begin{equation}\label{eq:sigS}
        \sigma_S^2\equiv\langle s_i^2\rangle=\sum_\ell\frac{2\ell+1}{4\pi}S_\ell,
      \end{equation}
      and thus we can write the bias of $\tilde{N}^f$ as:
      \begin{equation}\label{eq:noise_bias_bias}
        \langle \tilde{N}^f\rangle=\tilde{N}_\ell+\tilde{N}^w\,\sigma_S^2,
      \end{equation}
      where we have defined the combination
      \begin{equation}\label{eq:Nw}
        \tilde{N}^w\equiv\frac{1}{4\pi}\sum_iw_i^2.
      \end{equation}

    \subsubsection{Mask noise}\label{sssec:meth.disc.wnoise}
      Drawing intuition from our discussion of uncorrelated noise in the field, we can see that the pseudo-$C_\ell$ of the mask $\tilde{C}^w_\ell$ will also receive a white-noise-like contribution from the terms $i=j$ in Eq. \ref{eq:pcl_mask}. By inspection, we can see that this is none other than the quantity $\tilde{N}^w$ we just calculated. We will thus refer to $\tilde{N}^w$ as the ``mask shot noise''. In terms of it, we can thus write the mask power spectrum as
      \begin{equation}
        \tilde{C}^w_\ell=\tilde{S}^w_\ell+\tilde{N}^w,
      \end{equation}
      where the ``signal'' component is, mathematically:
      \begin{equation}
        \tilde{S}^w_\ell=\frac{1}{4\pi}\sum_{i\neq j}w_iw_j\,P_\ell(\mu_{ij}),
      \end{equation}
      with $\mu_{ij}\equiv\nv_i\cdot\nv_j$, and $P_\ell(\mu)$ the Legendre polynomials. In practice, the fastest way to estimate $\tilde{S}^w_\ell$ is not to compute the double sum above, but to compute $\tilde{C}^w_\ell$ from the harmonic transform of $w(\nv)$ (Eq. \ref{eq:sht_disc}), and then subtract $\tilde{N}^w$ from it, calculated using Eq. \ref{eq:Nw}.

      The presence of $\tilde{N}^w$ is a feature exclusive of the discrete origin of our observations. It sets a floor below which $\tilde{C}^w_\ell$ cannot deviate significantly and, for most common point distributions of interest in astrophysics, $\tilde{C}^w_\ell$ will, in fact, tend to $\tilde{N}^w$ at high $\ell$. This is contrary to the usual scenario for power spectrum estimation over incomplete skies. In more common cases, in which the mask is a smooth function, and has support over a significant fraction of the sky (compared to the angular scales of interest), $\tilde{C}^w_\ell$ decays with $\ell$. This causes the sum over $\ell''$ in Eq. \ref{eq:mcm_mask} to converge quickly, and the resulting mode-coupling matrix to be highly concentrated along the diagonal. It is therefore important to understand what the contribution from $\tilde{N}^w$ would be to $M^w_{\ell\ell'}$ and, in general to the ensemble average of $\tilde{C}^f_\ell$. Calling these contributions $M^{w,N}_{\ell\ell'}$, and $(\Delta\tilde{C}^f_\ell)_N$, we can calculate them by replacing $\tilde{C}^w_\ell$ with $\tilde{N}^w$ in Eq. \ref{eq:mcm_mask}, and then substituting the result into Eq. \ref{eq:pcl}. The resulting contribution to the mode-coupling matrix is:
      \begin{equation}\label{eq:mcm_N}
        M^{w,N}_{\ell\ell'}=\frac{2\ell'+1}{4\pi}\tilde{N}^w,
      \end{equation}
      where we have made use of one of the orthogonality relations of the Wigner-$3j$ coefficients (Eq. \ref{eq:w3j_orth2}). We can thus see that the mask shot noise leads to a contribution that couples all multipole $\ell$s with all other multipoles $\ell'$ (in principle up to $\ell'=\infty$). Although this does not imply that the mode-coupling matrix is necessarily singular, it does mean that an accurate implementation of the pseudo-$C_\ell$ estimator might involve (depending on the size of $\tilde{N}^w$), estimating $M^w_{\ell\ell'}$ up to very large values of both $\ell$ and $\ell'$ in order to account for mode coupling. This is problematic, since estimating $M^w_{\ell\ell'}$ is the most expensive step of the pseudo-$C_\ell$ estimator.

      An alternative to avoid this, is to estimate the contribution to $\langle \tilde{C}^f_\ell\rangle$ from the mask shot noise term, and to bring it to the left hand side of Eq. \ref{eq:pcl}. From our result above, this contribution is, simply:
      \begin{equation}
        (\Delta \tilde{C}^f_\ell)_N=\sum_{\ell'}M^{w,N}_{\ell\ell'}S_{\ell'}=\tilde{N}^w\sigma_S^2.
      \end{equation}
      Thus, we find that the contribution to the masked field's pseudo-$C_\ell$ coming from the mask shot noise is precisely equivalent to the unwanted contribution to our naive estimate of the noise bias in Eq. \ref{eq:noise_bias_bias}. This result allows us to construct a fully unbiased estimator for the power spectrum of a discretely-sampled field with uncorrelated noise, which we describe below.

    \subsubsection{An unbiased estimator}\label{sssec:meth.disc.estimator}
      Gathering our results from the last two sections, the algorithm presented in this paper is as follows:
      \begin{enumerate}
        \item Estimate the spherical harmonic coefficients of the masked field, and of the mask (Eq. \ref{eq:sht_disc}).
        \item From these, compute the pseudo-$C_\ell$ of the mask ($\tilde{C}^w_\ell$) and of the masked field ($\tilde{C}^f_\ell$).
        \item Estimate the ``noise'' components of both power spectra, $\tilde{N}^w$, $\tilde{N}^f$ directly from the measurements, using Eqs. \ref{eq:Nw} and \ref{eq:Nf}, respectively.
        \item Subtract both noise contributions to form the ``signal'' pseudo-$C_\ell$s of the mask and the masked field\footnote{As we discussed, $\tilde{S}^f_\ell$ is actually a biased estimator of the signal component of $\tilde{C}^f_\ell$, since we are subtracting a contribution sourced by the signal variance. Nevertheless, having corrected $\tilde{C}^w_\ell$ in the same way, the resulting estimator of $S_\ell$ will be unbiased.} $\tilde{S}^w_\ell\equiv\tilde{C}^w_\ell-\tilde{N}^w$, $\tilde{S}^f_\ell\equiv\tilde{C}^f_\ell-\tilde{N}^f$.
        \item Apply the standard pseudo-$C_\ell$ estimator to these ``signal-only'' spectra. This means, compute the mode-coupling matrix using $\tilde{S}^w_\ell$ instead of $\tilde{C}^w_\ell$, and apply its inverse\footnote{The usual caveats about power spectrum binning and invertible mode-coupling matrices should be borne in mind, however. See \cite{1809.09603}.} to $\tilde{S}^f_\ell$ to obtain an unbiased estimate of the signal power spectrum.
      \end{enumerate}
      To reiterate, subtracting the noise-like contribution from the mask pseudo-$C_\ell$ has two desirable consequences: first, it automatically allows us to account for the bias induced by correcting for a noise power spectrum computed from our own data. Secondly, the resulting mode-coupling matrix is now computed from the noise-debiased mask spectrum which, by construction, will decay with $\ell$ (at least for the usual point distributions that make up most cosmological observations). Hence, the sum over $\ell''$ in \ref{eq:mcm_mask} will have better convergence properties, and the mode-coupling matrix will be less numerically unstable.

    \subsubsection{Noise hardening}\label{sssec:meth.disc.hardening}
      The result presented in Section \ref{sssec:meth.disc.wnoise} may seem slightly disconcerting: by correcting for the noise-like component of the mask pseudo-$C_\ell$, which we can estimate entirely from the data, we are able to \emph{exactly} correct for the bias induced by the contribution to our estimate of the noise power spectrum coming from the signal variance. However, this contribution in principle depends on the true signal power spectrum, which we do not know a priori. For this to be possible, it must be that our correction of the mask power spectrum automatically makes the estimator immune to \emph{any} noise-like (i.e., constant additive) contribution to the power spectrum. In other words, the corrected mode-coupling matrix $M^{w,S}_{\ell\ell'}\equiv M^w_{\ell\ell'}-M^{w,N}_{\ell\ell'}$ must satisfy that $\sum_{\ell'} M^{w,S}_{\ell\ell'}S_{\ell'}=0$ if $S_\ell={\rm constant}$ for all $\ell$. In linear algebra terms, $M^{w,S}_{\ell\ell'}$ has a zero eigenvalue with eigenvector $v_\ell=1\,\forall\ell$. We can show this explicitly:
      \begin{align}\nonumber
        \sum_{\ell'}M^{w,S}_{\ell\ell'}
        &=\frac{1}{4\pi}\sum_{\ell'\ell''}(2\ell'+1)(2\ell''+1)\wtj{\ell}{\ell'}{\ell''}{0}{0}{0}^2[\tilde{C}^w_{\ell''}-\tilde{N}^w],\\\nonumber
        &=\sum_{\ell''}\frac{2\ell''+1}{4\pi}[\tilde{C}^w_{\ell''}-\tilde{N}^w],\\\nonumber
        &=\sum_{i\neq j}\frac{w_iw_j}{4\pi}\sum_{\ell'' m''}Y^*_{\ell'' m'',i}Y_{\ell'' m'',j},\\\nonumber
        &=\sum_{i\neq j}\frac{w_iw_j}{4\pi}\delta^D(\nv_i,\nv_j)=0,
      \end{align}
      where, in the second line, we have used the orthogonality relation Eq. \ref{eq:w3j_orth2}, in the third line we have expanded $\tilde{C}^w_\ell$ and $\tilde{N}^w$ using Eqs. \ref{eq:pcl_mask} and \ref{eq:Nw}, and in the last line we have used the completeness of the spherical harmonic functions (Eq. \ref{eq:Ylm_comp}).

      In practice, what this means is that this power spectrum estimator is not able to recover any completely flat component of the true power spectrum or, in other words, that it has infinite variance for these flat modes. This is not a strong shortcoming, since the amplitude of such modes can be reconstructed from the zero-lag correlation function (i.e., the variance of the field across all sources), assuming a model for the noise. This also means that, to avoid numerical instabilities, it will be useful to employ singular value decomposition methods before inverting the noise-corrected mode-coupling matrix. The crucial advantage is the fact that we do not need to know which fraction of the total field variance is attributed to uncorrelated noise and which fraction to the signal autocorrelation: changes in either of them will only affect, and are exactly captured by, $\tilde{N}^f$. This fundamentally establishes the unbiasedness of our estimator.

    \subsubsection{Cross-correlations}\label{sssec:meth.disc.cross}
      Cross-correlations between two different fields measured on two disjoint sets of points, or cross-correlations between standard maps and discretely-sampled fields, can be estimated by making use of the standard pseudo-$C_\ell$ algorithm, taking advantage of discrete harmonic transform methods in the case of the sampled fields. No additional care must be taken in this case about subtracting noise components in the mask or field power spectra, since these are zero by construction. Care must be taken, however, if the two different fields to be correlated have been sampled at the same discrete positions. A good example is the cross-correlation between the cosmic shear signal of a set of sources and the PSF ellipticity at their position, as a diagnostic for potential systematics. In the most general case, different weights may be applied to the sources when mapping either field. In this case, the mask and masked-field pseudo-$C_\ell$s must be corrected for noise bias.

      Let $e$ and $f$ be the two fields being correlated, and let $v_i$ and $w_i$, respectively, be the weights applied to each source when mapping either field. The noise-like contributions to the mask pseudo-$C_\ell$ and to the masked-field pseudo-$C_\ell$ are simple generalizations of those presented in the case of auto-correlations:
      \begin{equation}
        \tilde{N}^{vw}\equiv\frac{1}{4\pi}\sum_i v_i\,w_i,\hspace{12pt}\tilde{N}^{ef}\equiv\frac{1}{4\pi}\sum_i\,v_i\,w_i\,e_i\,f_i.
      \end{equation}

  \subsection{Spin-$s$ fields}\label{ssec:meth.spin}
    Let $f^a(\nv)$, with $a\in\{1,2\}$ be a spin-$s_f$ field, and let $f^\alpha_{\ell m}$ be its spherical harmonic coefficients (with $\alpha\in\{E,B\}$, corresponding to $E$- and $B$-modes). The two are related via a spin-$s_f$ spherical harmonic transform (and its inverse):
    \begin{equation}
      f^a(\nv)=\sum_{\ell m}\,_{s_f}Y^{a\alpha}_{\ell m}(\nv)f^\alpha_{\ell m},\hspace{6pt}\leftrightarrow f_{\ell m}^\alpha=\int d\nv\,f^a(\nv)\,_{s_f}Y^{a\alpha*}_{\ell m}(\nv),
    \end{equation}
    where the generalized spin-$s$ spherical harmonic functions $_sY^{a\alpha}_{\ell m}(\nv)$ are defined in Appendix \ref{app:shts}, and we implicitly sum over all repeated indices (e.g. $a$ and $\alpha$ above).

    The pseudo-$C_\ell$ of the masked field is defined in the same way as in the scalar case, except now we have to keep track of $E/B$-mode indices:
    \begin{equation}
      \tilde{C}^{f,\alpha\beta}_\ell\equiv\frac{1}{2\ell+1}\sum_m f_{\ell m}^\alpha f_{\ell m}^{\beta*}.
    \end{equation}
    The generalization of the scalar pseudo-$C_\ell$ estimator from Eq. \ref{eq:pcl} to spin-$s$ fields is:
    \begin{equation}\label{eq:pcl_spin}
      \left\langle\tilde{C}^{f,\alpha\beta}_\ell\right\rangle=\sum_{\ell'}({\sf M}^w_{\ell\ell'})^{\alpha\beta}_{\alpha'\beta'}\,S^{\alpha'\beta'}_{\ell'}+\tilde{N}^{\alpha\beta}_\ell,
    \end{equation}
    where the mode-coupling matrix is \cite{astro-ph/0410394,1809.09603}:
    \begin{equation}\label{eq:mcm_spin}
      {\sf M}_{\ell\ell'}^w=\frac{2\ell'+1}{4\pi}\sum_{\ell''}\tilde{C}^w_{\ell''}\,(2\ell''+1)\wtj{\ell}{\ell'}{\ell''}{s_f}{-s_f}{0}^2\,{\sf D}_{\ell+\ell'+\ell''},
    \end{equation}
    and the matrix ${\sf D}_n$ is:
    \begin{align}\label{eq:Dmat}
      &({\sf D}_n)^{EE}_{EE}=({\sf D}_n)^{EB}_{EB}=({\sf D}_n)^{BE}_{BE}=({\sf D}_n)^{BB}_{BB}=\frac{1+(-1)^n}{2},\\
      &({\sf D}_n)^{EE}_{BB}=-({\sf D}_n)^{EB}_{BE}=-({\sf D}_n)^{BE}_{EB}=({\sf D}_n)^{BB}_{EE}=\frac{1-(-1)^n}{2}.
    \end{align}

    Following the same logic used in Section \ref{ssec:meth.disc}, in the case of discretely-sampled fields, the spherical harmonic transform of a sampled and masked spin-$s_f$ field is
    \begin{equation}
      f^{w,\alpha}_{\ell m}=\sum_iw_if_i^a\,_sY^{a\alpha*}_{\ell m,i},
    \end{equation}
    and its pseudo-$C_\ell$ is
    \begin{equation}
      \tilde{C}^{f,\alpha\beta}_\ell=\sum_{ij}w_iw_jf_i^af_j^b\frac{1}{2\ell+1}\sum_m\,_sY^{a\alpha*}_{\ell m,i}\,_sY^{b\beta}_{\ell m, j}.
    \end{equation}

    Now, let us separate the field into signal and noise components $f^a=s^a+n^a$, where the signal is statistically isotropic ($\langle s^\alpha_{\ell m}s^{\beta*}_{\ell'm'}\rangle=\delta_{\ell\ell'}\delta_{mm'}S^{\alpha\beta}_\ell$), and the noise is uncorrelated between different sampling points, which in this case means $\langle n_i^an_j^b\rangle=\delta^{ab}\delta_{ij}\sigma^2_{N,i}$. Inserting this in the previous equation, and using Eq. \ref{eq:sYlm_add}, we obtain the noise bias:
    \begin{equation}
      \tilde{N}^{\alpha\beta}_{\ell}=\sum_iw_i^2\sigma_{N,i}^2\frac{1}{2\ell+1}\sum_{m}\,_sY^{a\alpha*}_{\ell m,i}\,_sY^{a\beta}_{\ell m,i}=\frac{\delta^{\alpha\beta}}{4\pi}\sum_iw_i^2\sigma_{N,i}^2.
    \end{equation}
    Based on this result, a potentially useful estimate of the noise bias, using only our measurements of $f$, would be:
    \begin{equation}\label{eq:Nf_spin}
      \tilde{N}^{f,\alpha\beta}\equiv\frac{\delta^{\alpha\beta}}{4\pi}\sum_i w_i^2\,\frac{1}{2}f_i^af_i^b\delta_{ab}.
    \end{equation}
    As in the spin-0 case, this estimator will be biased due to the contribution from the signal component of $f$. To calculate this bias, we take the expectation value of the previous equation, and find:
    \begin{align}\nonumber
      \langle \tilde{N}^{f,\alpha\beta}\rangle
      &=\tilde{N}_\ell^{\alpha\beta}+\frac{\delta^{\alpha\beta}}{4\pi}\sum_iw_i^2\frac{\delta_{ab}}{2}\langle s^a_is^b_i\rangle\\\nonumber
      &=\tilde{N}_\ell^{\alpha\beta}+\frac{\delta^{\alpha\beta}}{4\pi}\sum_iw_i^2\frac{1}{2}\sum_{\ell}S^{\alpha'\beta'}_{\ell}\sum_{m}\,_sY^{a\alpha'*}_{\ell m, i}\,_sY^{a\beta'}_{\ell m, i}\\\nonumber
      &=\tilde{N}_\ell^{\alpha\beta}+\delta^{\alpha\beta}\tilde{N}^w\sum_\ell\frac{2\ell+1}{4\pi}\frac{S_\ell^{EE}+S_\ell^{BB}}{2}\\
      &\equiv\tilde{N}_\ell^{\alpha\beta}+\delta^{\alpha\beta}\tilde{N}^w\sigma_S^2,
    \end{align}
    in analogy with Eq. \ref{eq:noise_bias_bias}, after generalising the expression for the field variance $\sigma_S^2$ in Eq. \ref{eq:sigS}. In the second line above, we have simply expanded $s_i^a$ in its spherical harmonic coefficients, and used their statistical isotropy. We have then made use of Eq. \ref{eq:sYlm_add} in the third line.

    The final step in our derivation shows that the contribution from $\tilde{N}^w$ to Eq. \ref{eq:pcl_spin} cancels out the term $\propto\sigma_S^2$ in the last equation, thus yielding and unbiased power spectrum estimator. To begin with, let us replace $\tilde{C}^w_\ell$ with $\bar{N}^w$ in Eq. \ref{eq:mcm_spin} to calculate the mask noise contribution to the mode-coupling matrix, ${\sf M}^{w,N}_{\ell\ell'}$:
    \begin{align}
      {\sf M}^{w,N}_{\ell\ell'}=\tilde{N}^w\,\frac{2\ell'+1}{4\pi}\sum_{\ell''}(2\ell''+1)\wtj{\ell}{\ell'}{\ell''}{s_f}{-s_f}{0}^2\,{\sf D}_{\ell+\ell'+\ell''}.
    \end{align}
    To simplify this further, note that all elements of ${\sf D}_{\ell+\ell'+\ell''}$ either are zero, or take the form $[1\pm(-1)^{\ell+\ell'+\ell''}]/2$ (see Eq. \ref{eq:Dmat}). We can use the term $\propto(-1)^{\ell+\ell'+\ell''}$ to swap the sign of the lower spin indices of one of the two Wigner $3j$ symbols, using Eq. \ref{eq:w3j_flip}. The resulting sum over $\ell''$ for this term is then zero by virtue of Eq. \ref{eq:w3j_orth2}, and we are left only with the contribution from the first term in the ${\sf D}$ matrix elements, which is always $1/2$ (or zero). Again, using Eq. \ref{eq:w3j_orth2} on this term, we obtain the simplified result:
    \begin{equation}
      {\sf M}^{w,N}_{\ell\ell'}=\tilde{N}^w\frac{2\ell'+1}{4\pi}\,\frac{{\sf d}}{2},
    \end{equation}
    where the matrix ${\sf d}$ has elements:
    \begin{equation}
      ({\sf d})^{EE}_{EE}=({\sf d})^{EB}_{EB}=({\sf d})^{BE}_{BE}=({\sf d})^{BB}_{BB}=({\sf d})^{EE}_{BB}=({\sf d})^{BB}_{EE}=1, \hspace{12pt} ({\sf d})^{EB}_{BE}=({\sf d})^{BE}_{EB}=-1,
    \end{equation}
    and zero otherwise. Substituting this in Eq. \ref{eq:pcl_spin}, and after a bit of algebra\footnote{This can be done by writing out ${\sf d}$ as a $4\times4$ matrix, and multiplying it by the vector $(S_\ell^{EE},S_\ell^{EB},S_\ell^{BE},S_\ell^{BB})$, with $S_\ell^{BE}=S_\ell^{EB}$.}, we find that the resulting contribution to $\langle\tilde{C}^{f,\alpha\beta}_\ell\rangle$ is
    \begin{equation}
      (\Delta\tilde{C}^{f,\alpha\beta}_\ell)_N=\delta^{\alpha\beta}\tilde{N}^w\sigma_S^2.
    \end{equation}

    Thus, the recipe for an unbiased estimator described in Section \ref{sssec:meth.disc.estimator} can be directly applied to spin-$s$ fields, simply replacing the scalar noise bias $\tilde{N}^f$ by its spin generalization $\tilde{N}^{f,\alpha\beta}$ in Eq. \ref{eq:Nf_spin}.

    \subsubsection{Different spins}\label{sssec:meth.spin.diff}
      One important generalization of the formalism we just described is the calculation of cross-power spectra between two fields of different spin sampled at the same source positions. A good example of this would be measurements of both the shape (spin-2) and size (spin-0) of galaxies, as proxies for weak lensing shear and convergence. The formalism we just described, in which the mode-coupling matrix is estimated from the noise-debiased mask pseudo-$C_\ell$, can be applied directly to these situations without any modifications. In fact, the orthogonality relation Eq. \ref{eq:w3j_orth2} guarantees that, for fields of different spin, the contribution from $\tilde{N}^w$ to the mode-coupling matrix vanishes exactly. Likewise, by virtue of the addition theorem for spin-weighted spherical harmonics (Eq. \ref{eq:sYlm_add}), the noise bias to the pseudo-$C_\ell$ of the masked fields also vanishes in the limit of uncorrelated noise. Thus, in these cases, one may even forego the calculation of $\tilde{N}^w$ and $\tilde{N}^f$ altogether.

  \subsection{Clustering}\label{ssec:meth.clust}
    A conceptually different type of discretely sampled field is the case of galaxy clustering, in which the overdensity of points in the discrete source catalog is itself the field. This case has been discussed in detail in \cite{2312.12285}, but we repeat part of the discussion here for completeness. In this case, the number density of objects is
    \begin{equation}
      n_d(\nv)=\sum_{i=1}^N w^d_i\,\delta^D(\nv,\nv_i),
    \end{equation}
    where $w^d_i$ is the weight of the $i$-th galaxy (which may be used to account for observational conditions, instrumental effects, such as fiber collisions, and other systematics \cite{2011.03408}).

    Critical in the study of galaxy clustering is our ability to account for spatial variations in the expected mean number density of sources $\bar{n}(\nv)$, caused by depth variations across the footprint, contamination by galactic or instrumental systematics, and the complex geometry of the observed footprint itself. This has been commonly done through the use of ``random catalogs'': synthetic catalogs containing discrete points that trace the underlying expected density of sources, but that are otherwise not clustered. The mean number density of sources in this case is then
    \begin{equation}
      \bar{n}(\nv)=\alpha\sum_{i=1}^{N_r} w_i^r\,\delta^D(\nv,\nv_i),
    \end{equation}
    where $w_i^r$ is the weight of the $i$-th random point, $N_r$ is the total number of random points, and the normalization factor $\alpha$ accounts for the potentially different number of objects in the data and random catalogs:
    \begin{equation}
      \alpha\equiv\frac{\sum_{i=1}^Nw_i^d}{\sum_{i=1}^{N_r}w_i^r}.
    \end{equation}
    Ideally, the field whose power spectrum we wish to estimate is the galaxy overdensity $f(\nv)\equiv\delta_g(\nv)\equiv n_d(\nv)/\bar{n}(\nv)-1$, although, obviously, the discretized nature of $\bar{n}(\nv)$ complicates the construction of such a map. We can proceed, however, under the following consideration: a close to the optimal choice of mask for the standard pseudo-$C_\ell$ method is one that is proportional to the inverse noise variance of the field under study. Assuming $n_d$ to be a Poisson realization of $\bar{n}(1+\delta)$, a reasonable choice of mask is therefore $w(\nv)=\bar{n}(\nv)$. With this choice, the masked field and the associated mask are simply
    \begin{align}
      &f^w(\nv)=n_d(\nv)-\bar{n}(\nv)=\sum_{i=1}^Nw_i^d\,\delta^D(\nv,\nv_i)-\alpha\sum_{i=1}^{N_r}w_i^r\,\delta^D(\nv,\nv_i),\\
      &w(\nv)=\alpha \sum_{i=1}^{N_r} w_i^r\delta^D(\nv,\nv_i).
    \end{align}

    The methodology described in Section \ref{sssec:meth.disc.estimator} can therefore be extended directly to the case of galaxy clustering with the following changes:
    \begin{itemize}
      \item The SHTs of the masked field and the mask are
      \begin{equation}\label{eq:sht_gc}
        f^w_{\ell m}=\sum_{i=1}^Nw_i^dY^*_{\ell m,i}-w_{\ell m},\hspace{12pt} w_{\ell m}=\alpha\sum_{i=1}^{N_r}w_i^rY^*_{\ell m,i},
      \end{equation}
      \item The shot noise components to field and mask are:
      \begin{equation}
        \tilde{N}^f=\frac{1}{4\pi}\sum_{i=1}^N(w_i^d)^2+\tilde{N}^w,
        \hspace{12pt}
        \tilde{N}^w=\frac{1}{4\pi}\sum_{i=1}^{N_r}(\alpha w_i^r)^2.
      \end{equation}
    \end{itemize}

    Although the use of randoms to characterize survey homogeneity has been common, particularly in the analysis of spectroscopic galaxy surveys, it is also possible to map spatial variations in survey homogeneity at every point in the celestial sphere. It is therefore important to also consider the case in which a ``completeness map'' $W$ is provided, such that $W(\nv)\propto\bar{n}(\nv)$. In this case, the method must be adapted as follows:
    \begin{itemize}
      \item The sky mask is generated by normalizing the completeness map, to make it correspond to the expected number density of galaxies:
      \begin{equation}
        w(\nv)=\alpha W(\nv),\hspace{12pt}\alpha=\frac{\sum_{i=1}^Nw_i^d}{\int d\nv\,W(\nv)}.
      \end{equation}
      \item The SHT of the masked field is still given by the first equality in Eq. \ref{eq:sht_gc}, with the mask SHT $w_{\ell m}$ estimated from the continuous $w(\nv)$ as usual.
      \item Since the mask is now a map, rather than a collection of sources, its shot noise is $\tilde{N}^w=0$. The shot-noise component of the field's pseudo-$C_\ell$, in turn, is $\tilde{N}^f=\sum_{i=1}^N (w_i^d)^2/(4\pi)$.
    \end{itemize}

\section{Validation} \label{sec:val}
  Having presented the mathematical formalism of an unbiased power spectrum estimator for discretely sampled fields, we now proceed to validate it against simulations. We focus on what is likely the most relevant probe for which this formalism is useful: cosmic shear catalogs. The results presented here will therefore correspond to spin-2 fields, but other spin cases do not change the problem conceptually\footnote{Except perhaps in the case $s=0$, where the separation between  $E$- and $B$-modes is meaningless.}. We showcase examples of the spin-1 and spin-0 cases in Sections \ref{ssec:ex.quaia} and \ref{ssec:ex.frb}, respectively.

  Cosmic shear measures the apparent distortion of a galaxy's shape due to the gravitational lensing effect caused by the foreground matter distribution integrated along the line of sight. This effect, while weak and indistinguishable from the galaxy's intrinsic ellipticity when considering single objects, represents a powerful probe of the cosmic large-scale structure when studied in a statistical way on large catalogs of galaxies. In the weak lensing regime, the shear $\gamma_{a,i}$ that the $i$-th galaxy's image has undergone, provides a linear contribution to its measured ellipticity $e_{a,i}$ (estimated from the traceless component of the projected inertia tensor, see e.g. \cite{2010.09717}):
  \begin{equation}
    e_{a,i} = \gamma_{a,i} + n_{a,i},
  \end{equation}
  where $a=1,2$ labels the two components of ellipticity and shear, and the noise component $n_{a,i}$ includes both the intrinsic (i.e., unlensed) galaxy ellipticity and any additional measurement noise. Crucially, it is common to work under the assumption that all contributions to $n_{a,i}$ are uncorrelated between different sources.

  Section \ref{ssec:val.setup} below describes the setup used to validate the catalog-based methodology described in the previous section, comparing it against the results of the standard map-based approach. The results of this validation, as a function of the number density of catalog sources, are then presented in Section \ref{ssec:val.val}.

  \subsection{Validation setup}\label{ssec:val.setup}
    To validate our estimator, we generate 500 synthetic cosmic shear catalogs. Each catalog is created by first generating a sky realization of the spin-2 cosmic shear field as a Gaussian random field from a theoretical $E$-mode power spectrum. This realization is generated using the \hp pixelization scheme with a resolution parameter $N^{\rm hires}_{\rm side}=4096$, corresponding to a pixel size of $\sim0.86'$. The native resolution is chosen to be significantly higher than the angular scales we will try to extract using either the map-based or catalog-based methods, to ensure that the finite-resolution effects inherent to these ground-truth maps are negligible compared to the typical statistical error. From this map, we then produce a mock observed shear catalog, by sampling it at the positions of a discrete set of sources. Specifically, to each source we assign the value of the shear field corresponding to the \hp pixel it falls in\footnote{Unfortunately, more sophisticated interpolation schemes are highly computationally intensive for \hp.}. We do not add any shape noise to the catalog, in order to reduce the statistical uncertainties of the recovered power spectra, thereby maximizing our sensitivity to any potential bias.

    Once a mock shear catalog has been generated, we estimate its angular power spectrum using the catalog-based method described above, and the standard map-based method. For the map-based case, we follow the procedure outlined in \cite{2010.09717}: a map of the average shear in each pixel is constructed at a resolution $N_{\rm side}=1024$. In each pixel, the local shear field is calculated as the weighted average of the source ellipticities in it. As our simulated catalogs do not contain any additional shape noise, we do not subtract any data-based noise bias in the standard map-based approach (although we do subtract $\tilde{N}^f$ and $\tilde{N}^w$ in the catalog-based method, as described in Section \ref{sec:meth}). The \hp pixelization suppresses the small-scale information in the shear map, reducing the map RMS by a pixel window function, which needs to be accounted for when computing pseudo-$C_\ell$s \cite{1004.3542}. Assuming that the pixelization is coarse enough so that every pixel contains at least one galaxy, we can approximate the pixel window function in real space by a Gaussian smoothing kernel with an FWHM of $\sim 41.7$ deg$/N_{\rm side}$ size \cite{astro-ph/0409513}. We do not correct for this bias in the simulated data, but account for it when comparing with the bandpower-convolved theory power spectrum.

    In order to produce mock catalogs with realistic source distributions, we use the angular positions and shape weights of real data from the public KiDS-1000 shear catalog \cite{2007.01845}. In particular, we use all sources from the fourth photometric bin, which contains $4{,}544{,}395$ sources, observed in two different fields, covering an area of approximately $800\,{\rm deg}^2$ after masking. The ground-truth maps were generated using the theoretical prediction for this redshift bin based on the best-fit $\Lambda$CDM parameters from \planck \cite{1807.06209}, assuming no intrinsic alignments. We note that, by choosing to use real KiDS-1000 sources, we ensure that the method is validated on catalogs with not only a realistic number density, but also a spatial distribution that reproduces both the expected clustering of sources, and their inhomogeneity caused by depth variations across the observed footprint. In order to quantify the performance of the catalog- and map-based methods as a function of source number density, we repeat our analysis on down-sampled versions of these catalogs, in which we select a fraction of the KiDS-1000 sources. Specifically, we will study the results of selecting only 1\% and 0.01\% of the sources. We choose a multipole binning of constant width $\Delta\ell=30$ up to a maximum of $\ell_{\rm max}=2N_{\rm side}=2048$.

  \subsection{Validation results}\label{ssec:val.val}

    \begin{figure*}
    \centering
     \includegraphics[width=\textwidth]{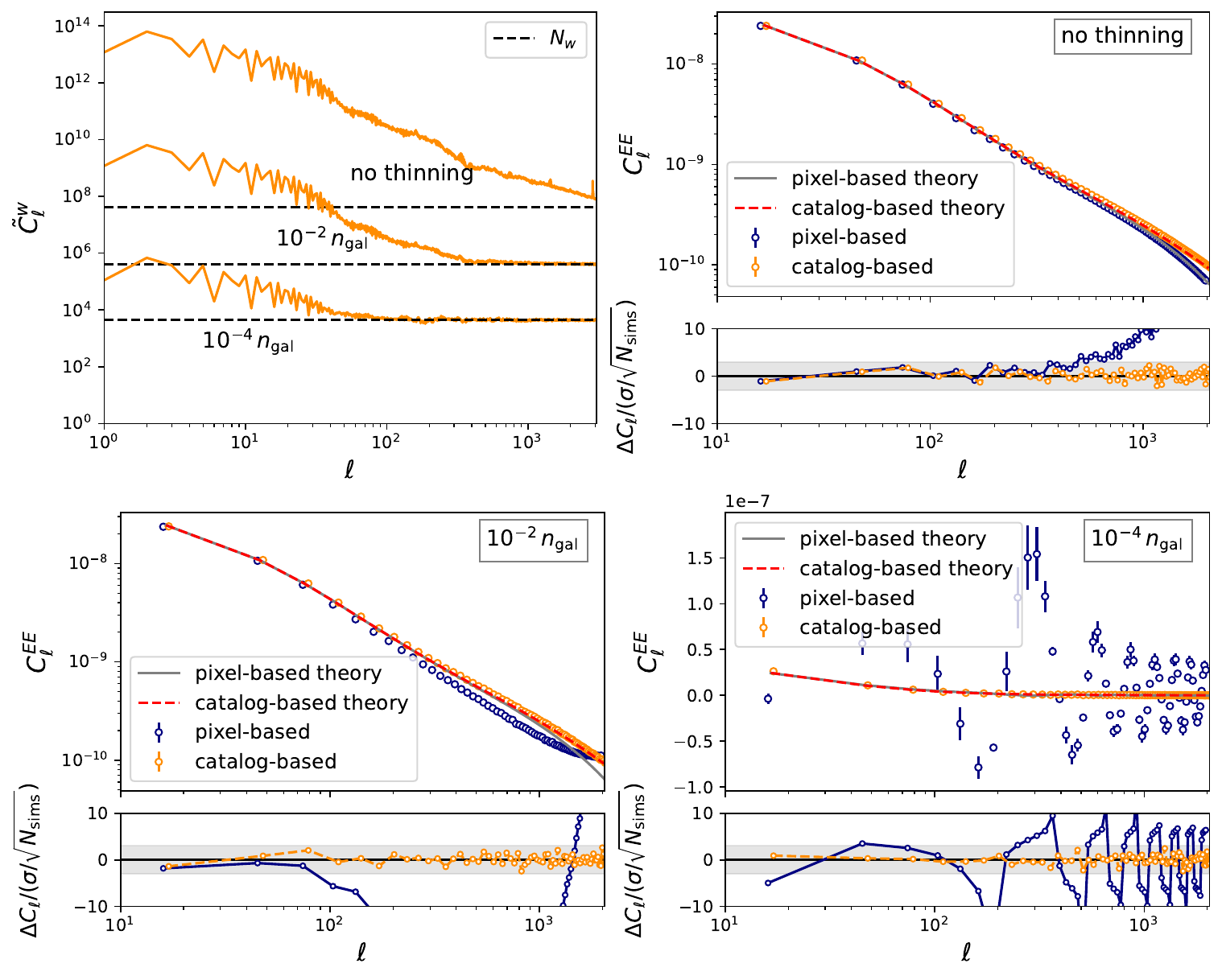}
    \caption{Validation shear $EE$ power spectra from 500 KiDS-1000-like mock catalogs with different thinning factors. \textit{Left upper panel:} Coupled pseudo power spectrum of the sky mask using the catalog-based approach (orange dashed line). Black horizontal dashed lines mark the mask shot noise level, which we subtract before computing the mode coupling matrix. The three sets of lines from top to bottom are the KiDS-1000 catalog without thinning, with a thinning factor of 100, and a thinning factor of 10,000. \textit{Right upper panel:} Mean decoupled power spectrum from simulations, evaluated with the pixel-based (blue circles) and the catalog-based approach (orange circles, including a horizontal offset for clarity), compared to theory bandpowers computed using either the pixel-based or the catalog-based mode coupling (grey solid lines and red dashed lines, respectively). The narrow subpanel below shows residuals between estimates and theory (corrected for the corresponding \hp pixel window), measured in units of the expected error on the mean. \textit{Lower panels:} results from 500 simulated catalogs that contain 100 times and $10^4$ times less sources than the original catalog.} \label{fig:validation}
    \end{figure*}

    Figure~\ref{fig:validation} shows the main results of our validation exercise. The upper right panel shows the results in the fiducial setup, in which all the KiDS-1000 sources are used to measure the power spectrum. The power spectrum averaged over all simulations is shown as orange and blue circles for the catalog- and map-based approaches, respectively. The theoretical predictions, estimated by convolving the input power spectrum with the bandpower window functions corresponding to the mode-coupling matrix derived with either method, are shown as dashed red and solid grey lines (the latter barely visible behind the red dashed lines). We multipliy the theory spectra by the squared \hp pixel window function ($N_{\rm side}^{\rm hires}=4096$ for the catalog-based estimator, $N_{\rm side}=1024$ for the map-based one). The narrow subpanel below shows the residual of each method with respect to its corresponding prediction normalized by the error on the mean of the 500 simulations (i.e., the scatter across simulations divided by $\sqrt{500}$). The horizontal grey bands show 3-$\sigma$ confidence intervals.
    
    The catalog-based method is able to produce an unbiased estimate of the power spectrum, while at $\ell\gtrsim 500$, the map-based approach deviates significantly due to pixelization effects at the map level and shot-noise-related bias in the estimator itself. Note that, in a realistic scenario, the statistical uncertainties would be at least $\sqrt{500}\sim22$ times larger, not accounting for the impact of shape noise, and hence these deviations would likely be subdominant. We discuss the various sources of this bias in Appendix \ref{app:bias}. For the catalog-based estimator, we must account for the pixelization of our ground truth shear map, and can do so accurately by means of the \hp window function.\footnote{The simulated catalog contains field values that are, by design, constant within every pixel of the ground truth map. This makes the catalog-based field effectively band-limited, with no power at scales below the pixel scale, and we can account for this by means of the analytic pixel window function.} Ideally, we could choose an even higher $N_{\rm side}^{\rm hires}$ to avoid this effect, but given the higher numerical cost, we prefer to correct the theory accordingly. With the map-based approach, the estimate is biased both for the uncorrected case (not shown in the figure) and the case where the theory is convolved with the pixel window function (shown). As explained in Appendix \ref{app:bias}, the two main reasons are the impact of aliasing, and the numerical instability of the pixel-based mode-coupling matrix due to mask shot noise. Note that this coupling matrix is truncated to multipoles $\ell<3N_{\rm side}$, and thus fails to capture the full correlation between scales induced by the mask for the sparsest catalogs.

    The bottom panels of Fig. \ref{fig:validation} show the results of repeating this analysis on catalogs containing 100 times and 10{,}000 times fewer sources (left and right panels, respectively). We find that, in spite of the substantial drop in number density, the catalog-based method is able to recover unbiased power spectrum measurements in both cases within the range of scales of interest. The map-based approach, however, recovers significantly biased measurements, which cannot be corrected for by our analytical pixel window function. This bias becomes gradually larger on smaller scales for the $1\%$ catalog, reaching very significant levels after $\ell\sim100$. For the $0.01\%$ catalog, in turn, the map-based method fails catastrophically, due to significant shot noise in the field pseudo-$C_\ell$ estimator, as well as numerical instabilities in the estimated mode-coupling matrix. These lead to the method misestimating the measured power spectrum at all scales (bottom right panel). In contrast, the catalog-based approach remains unbiased even for this extremely low number density (the shear field is sampled at the positions of just 454 sources). For the thinned catalogs, the catalog-based estimator is again fully unbiased even at scales $\ell>1024$ after accounting for pixelization-related bias mentioned above, which becomes less important for smaller source densities due to higher estimator noise ($\sim2\sigma$ and $\lesssim1 \sigma$ at $\ell=1024$ for 1\% and 0.01\% thinning, respectively).

    The upper left panel of Fig. \ref{fig:validation} shows the mask pseudo-$C_\ell$ in the catalog-based case, $\tilde{C}^w_\ell$, for the fiducial catalog, and the 1\% and 0.01\% catalogs (from top to bottom). In each case, the mask shot noise level $\tilde{N}^w$ is marked by horizontal black dashed lines, and the raw mask pseudo-$C_\ell$s are shown as the solid lines. We see that, for the original KiDS source density, the mask shot noise is low enough to not affect the mode-coupling matrix significantly, except above $\ell\sim2000$. In turn, the 1\% catalog becomes shot-noise dominated at $\ell\sim200$ (compared to $\ell\sim100$ where the decoupled power spectrum estimate starts deviating significantly from the input). Finally, the $0.01\%$ catalog is shot-noise dominated at almost all scales, leading to severe numerical instabilities in the standard pixel-based algorithm, which does not account for the impact of mask shot noise.

    \begin{figure*}
    \centering
     \includegraphics[width=\textwidth]{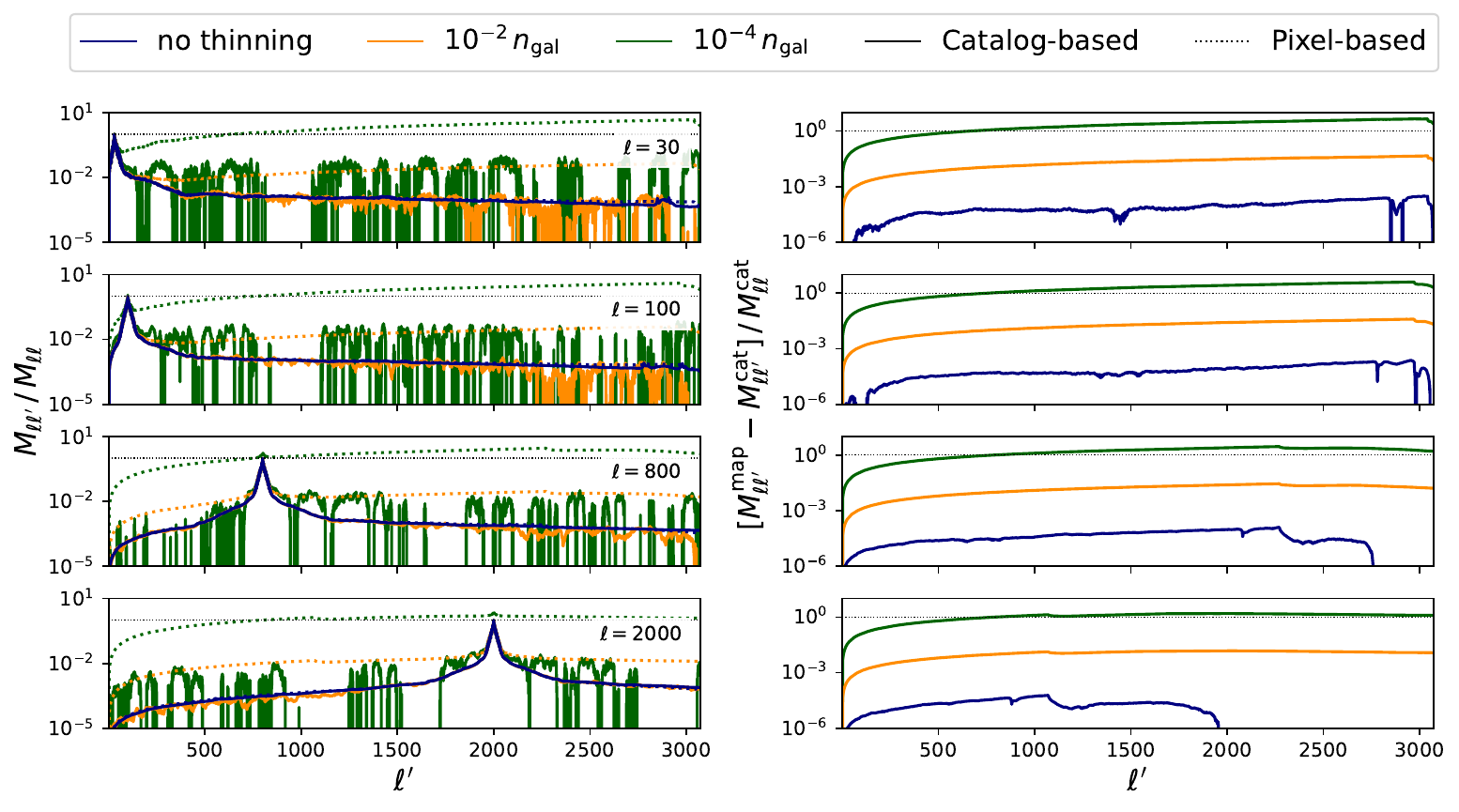}
    \caption{Rows of catalog-based mode coupling matrix and comparison between pixel-based and catalog-based approach, displayed for KiDS-1000-like mock catalogs of three different thinning factors. Blue, orange, and green are the cases of no thinning, thinning factor 100, and thinning factor 10,000. From top bottom, we show different rows of the mode coupling matrix: $\ell=30$, 100, 800, 2000. \textit{Left panels:} Mode coupling matrix for catalog- (solid lines) and pixel-based $C_\ell$s (dotted lines), normalized to the value of the $\ell=\ell'$ element. \textit{Right panels:} Difference between the rows of the mode coupling matrix in the map-based and the catalog-based case, normalized by the catalog-based element at $\ell=\ell'$.} \label{fig:mcm}
    \end{figure*}

    Figure~\ref{fig:mcm} shows the mode coupling matrices computed from mock catalogs, and compares the result to the map-based analog. We show this for the full KiDS-1000 catalog (blue) and the two thinned versions (orange, green) but focus on the former for now. The left panels show specific columns of the ${}^{EE}_{EE}$ component of the catalog-based corrected mode coupling matrix $M_{\ell\ell'}^{w,S}$ (solid lines, see Sect.~\ref{sssec:meth.disc.hardening}) and the standard pixel-based mode coupling (dashed lines), normalized to one at $\ell=\ell'$. The right panels show, in the same units, the relative difference between the columns of the map-based mode coupling matrix \eqref{eq:mcm_spin} and the catalog-based one\footnote{For a direct comparison between the catalog-based and the map-based mode coupling matrix elements, we need to divide the latter by the squared pixel area, $(4\pi/N_{\rm pix})^2$.}. From top to bottom, we display columns with $\ell=(30,\, 100,\, 800, \, 2000)$. We find that, in the case of the full catalog, both matrices differ by a factor of less than $10^{-4}$, indicating that the mask shot noise correction is subdominant. The matrix columns peak at $\ell=\ell'$ as expected, rapidly decline to percent level at $|\ell-\ell'|\sim 200$, and converge to a nearly constant value at the $0.1$ percent level. 

    For the $1\%$-thinned catalog, both methods start differing at the percent level. The elements of the catalog-based matrix match on average those for the full catalog, but are much noisier for couplings between distant multipoles, while the pixel-based coupling matrix includes a now very significant contribution due to mask shot noise, with the expected scaling $\propto (2\ell+1)$, see Eq.~\eqref{eq:mcm_N}. For the highly shot-noise dominated case with 454 sources, the catalog-based mode coupling becomes more noisy, with fluctuations reaching percent level, but its mean stays compatible with the one for the full catalog. On the other hand, the pixel-based coupling matrix is almost completely dominated by the shot noise component, and differs from the catalog-based version by more than 100\% at $\ell\gtrsim500$. This shows that even with an extremely low source number density, the catalog-based approach recovers a stable mode coupling matrix that allows us to estimate unbiased power spectra from sparsely sampled sources on the sky. 
    
    \begin{figure*}
    \centering
     \includegraphics[width=\textwidth]{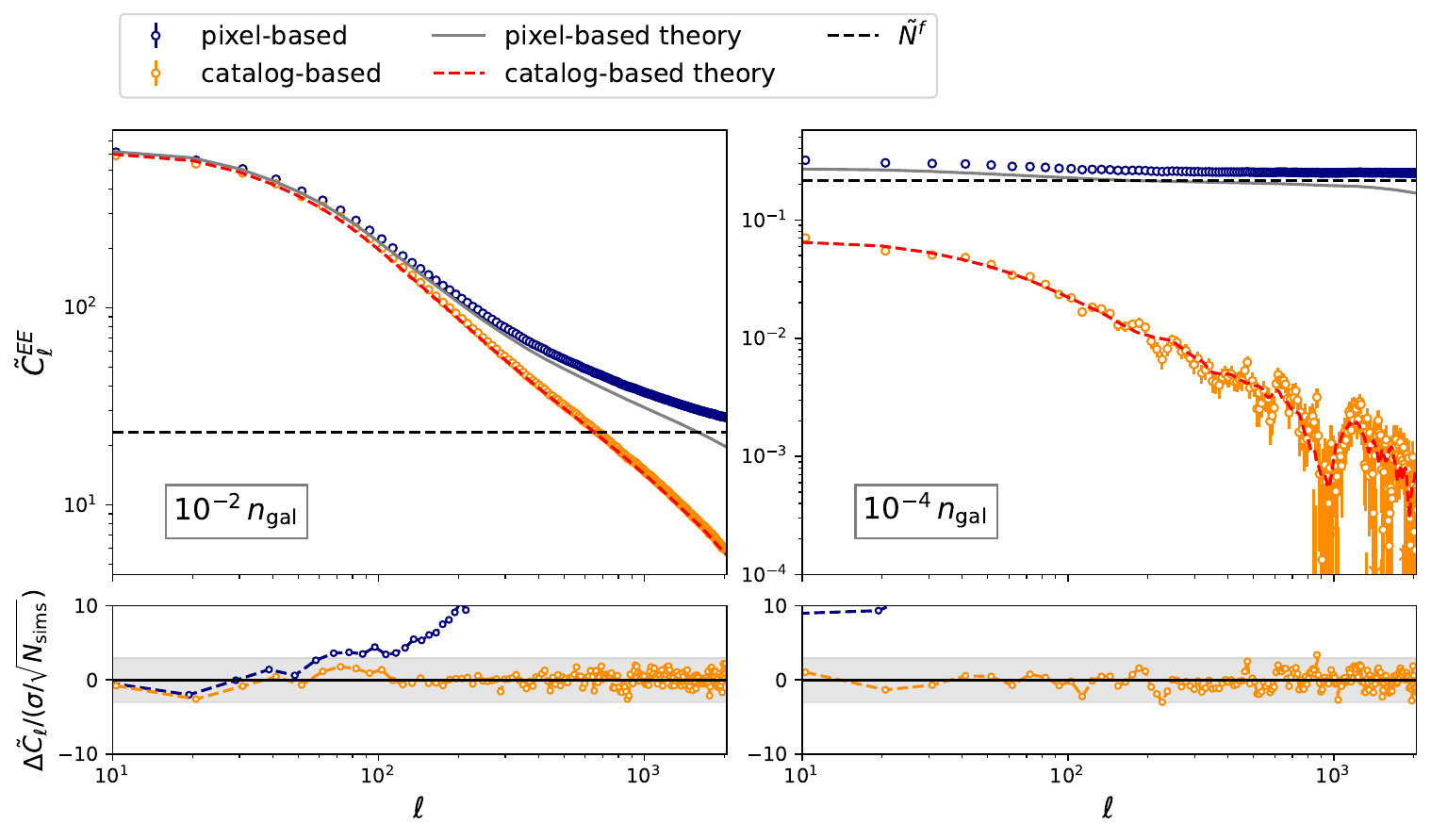}
    \caption{Coupled pseudo power spectra for KiDS1000-like validation catalogs, thinned by factors 100 and 10,000. The blue and orange error bars are centered at the mean from 500 simulations and their length corresponds to the standard deviations across 500 simulations divided by $\sqrt{500}$ in the map-based and catalog-based approach, respectively. The lower panels show the deviation between data and coupled theory, normalized by the error. The difference between both estimates corresponds to the total field noise $N_f$ (black dashed line, see Eq.~\ref{eq:noise_bias_bias}), which in this example is purely due to shot noise.} \label{fig:validation_coupled}
    \end{figure*}

    To understand what causes the bias in the pixel-based estimator, we compare both approaches in terms of the coupled pseudo-$C_\ell$ estimators. Figure~\ref{fig:validation_coupled} shows, in the same color scheme as in Fig.~\ref{fig:validation}, the coupled shear $EE$ pseudo-$C_\ell$s (see Eq.~\ref{eq:pcl_spin}) computed using the catalog-based and pixel-based approach, and compared to the input theory, multiplied by the square of the respective \hp pixel window, and coupled with the respective mode coupling matrix. The field noise level $N_f$ is shown as a horizontal black dashed line for comparison. The left and the right panel show the cases of $1\%$ and $0.01\%$ thinning, respectively, and the lower panels show the bias with respect to the mode-coupled theory prediction. The pixel- and catalog-based coupled estimators differ at similar scales as their decoupled versions shown in Fig.~\ref{fig:validation}, but without being affected by the numerical instability of the inverse binnned mode-coupling matrix.
    
    In contrast, the coupled theory bandpowers do depend on the mode coupling matrix. We see that the catalog-based theory prediction becomes noisy for small scales, which is inherited from the noisy mode-coupling matrix seen in Fig.~\ref{fig:mcm}. For both thinned catalogs, the catalog-based estimator is unbiased, while the map-based estimator receives a bias at intermediate and small scales. In summary, Fig.~\ref{fig:validation_coupled} illustrates that for low source number density, the pixel-based coupled pseudo power spectrum is shot-noise dominated. Analogously, the mask shot noise dominates the pixel-based mode coupling at small scales and leads to a severe misestimation of the mode-coupled theory theory prediction. The decoupled pixel-based estimator inherits this field noise bias, and suffers from numerical instabilities from applying the inverse mode coupling matrix. In contrast, the catalog-based estimator does not deviate significantly from theory predictions, even when considering the statistical power of 500 noiseless simulations combined.

\section{Example applications} \label{sec:ex}
  \subsection{Spin-2 example: cosmic shear from DES-Y3} \label{ssec:ex.des}
  As a first real-data example, we apply our estimator to the highest redshift bin of the galaxy shear catalog from DES-Y3 \cite{2011.03408}. This catalog contains $\mathcal{O}(10)$ more galaxies in a four times larger survey footprint than the fourth redshift bin of KiDS-1000 used in Sect.~\ref{sec:val}. Cosmic shear is a common example of a spin-2 field sampled at discrete positions, and one of the most powerful probes of the cosmic LSS. It measures the lensing-induced distortion of galaxy shapes due to the foreground matter distribution along the line of sight. In the Limber approximation \citep{1953ApJ...117..134L}, the cosmic shear angular power spectrum is related to the 3D matter power spectrum $P(k, \, z(\chi))$, and can be modeled as
  \begin{equation}
    C_\ell = G_\ell^2 \int \frac{\dint \chi}{\chi^2} \, q^2(\chi) \, P\left( k=\frac{\ell + 1/2}{\chi}, \, z(\chi) \right) \, ,
  \end{equation}
  where $k$ is the modulus of the Fourier wave vector, $z$ is the redshift, and $\chi$ is the comoving radial distance. The lensing kernel $q(\chi)$ quantifies the radial contribution of foreground matter to the angular galaxy lensing power spectrum, and depends on the redshift distribution $p(z)$ of galaxies that may act as lensed images. In the case of a flat universe in which general relativity is valid and $c=1$, the lensing kernel is given by
  \begin{equation}
    q(\chi) = \frac{3}{2} H_0^2 \Omega_m \frac{\chi}{a(\chi)} \int_{z(\chi)}^{\infty} \dint z’ \, p(z’)\frac{\chi(z’) - \chi}{\chi(z’)} \, ,
  \end{equation}
  where $H_0$ is the Hubble constant, $\Omega_m$ is the matter density parameter, and $a=1/(1+z)$ is the scale factor. Transitioning from the 3D Laplacian of the gravitational potential to the angular Hessian of the associated lensing potential requires the correction factor \citep{1702.05301}
  \begin{equation}
    G_\ell \equiv \sqrt{\frac{(\ell+2)!}{(\ell-2)!}}\frac{1}{(\ell + 1/2)^2}.
  \end{equation}
  Galaxy shear data are notoriously biased by galaxy shape noise, which has to be corrected for explicitly in the pixel-based approach \cite{2010.09717}, but is removed at the catalog level using our estimator.

\begin{figure}
    \centering
    \includegraphics[width=0.36\textwidth]{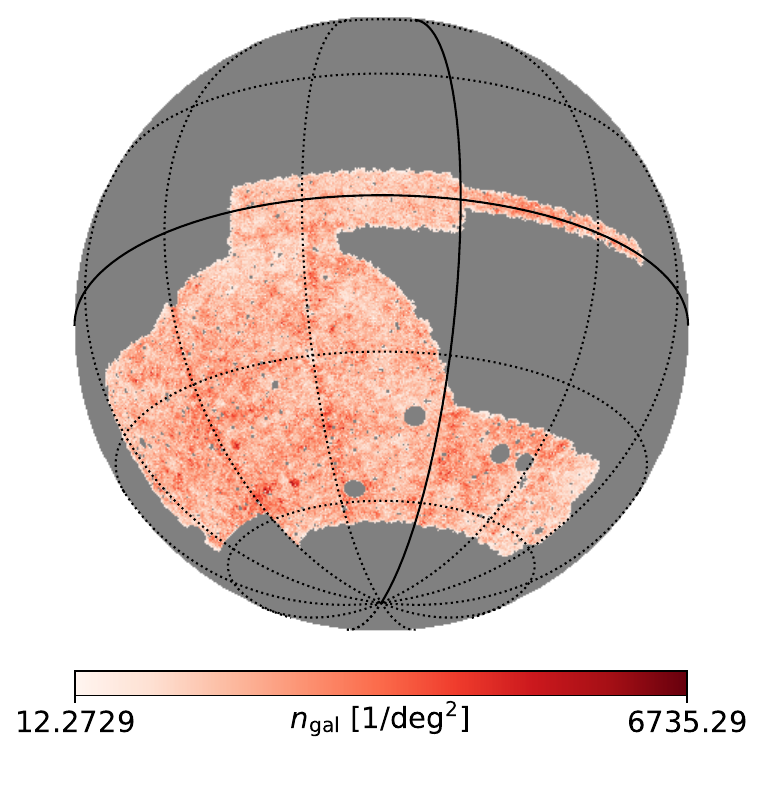}
    \includegraphics[width=0.6\textwidth]{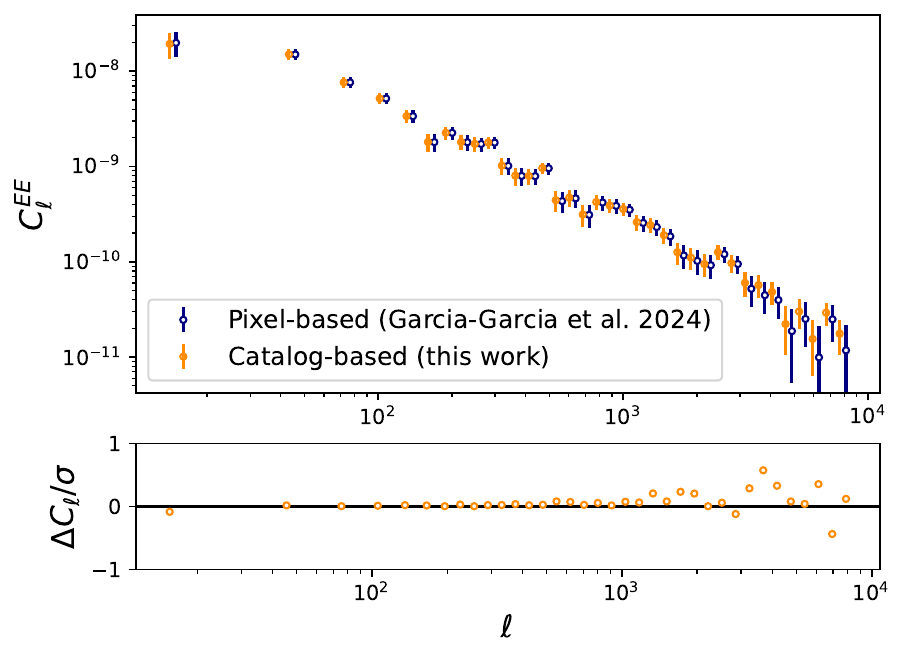}
    \caption{\textit{Left panel:} Galaxy source catalog of the fourth redshift bin of DESY3 data used in this work, displayed in celestial coordinates. \textit{Right panel:} decoupled $C_\ell^{EE}$ using the catalog-based approach (orange), and comparing with literature results from \cite{2403.13794} (blue, including a horizontal offset for clarity). We consider error bars from Gaussian and super-sample covariance terms taken from \cite{2403.13794}. The lower panel shows residuals of the catalog-based method with respect to the literature in units of the 1-sigma error.}
    \label{fig:des_data}
  \end{figure}

  We use the fourth redshift bin of the DES-Y3 data, containing 25,091,297 catalog galaxies between redshifts $\sim0.4$ and $\sim1.5$. We correct the catalog galaxy ellipticities for linear and multiplicative bias as described in \cite{2403.13794}. The survey footprint covers an effective sky area of 4143 deg$^2$ as shown in the left panel of Figure \ref{fig:des_data}. The weighted source number density of DES-Y3 is $n_{\rm eff} = 5.59\, {\rm gal}/{\rm arcmin}^2$, compared to $6.17\, {\rm gal}/{\rm arcmin}^2$ for KiDS-1000, yielding a similar mask shot noise level.

  Figure \ref{fig:des_data} in the right panel shows the results in terms of the decoupled power spectrum, comparing our catalog-based approach with recent pixel-based measurements using the pixel-based approach published in \cite{2403.13794}. Both estimators use the same set of multipole bins up to $\ell=8192$, corresponding to $2\,N_{\rm side}$ of the \hp map used with the pixel-based estimator, or an angular scale of $\sim26$ arcsec. We corrected the map-based estimate by the analytical pixel window function. Both results use 1-sigma error bars from \cite{2403.13794}, derived from an empirical power spectrum covariance that includes a Gaussian and a non-Gaussian super-sample contribution. The lower panel shows the relative difference between both methods in units of sigma. We find that both methods agree within 1 sigma within the shown multipole range.

  \subsection{Spin-1 example: proper motion surveys}\label{ssec:ex.quaia}
    \begin{figure}
        \centering
        \includegraphics[width=0.54\textwidth]{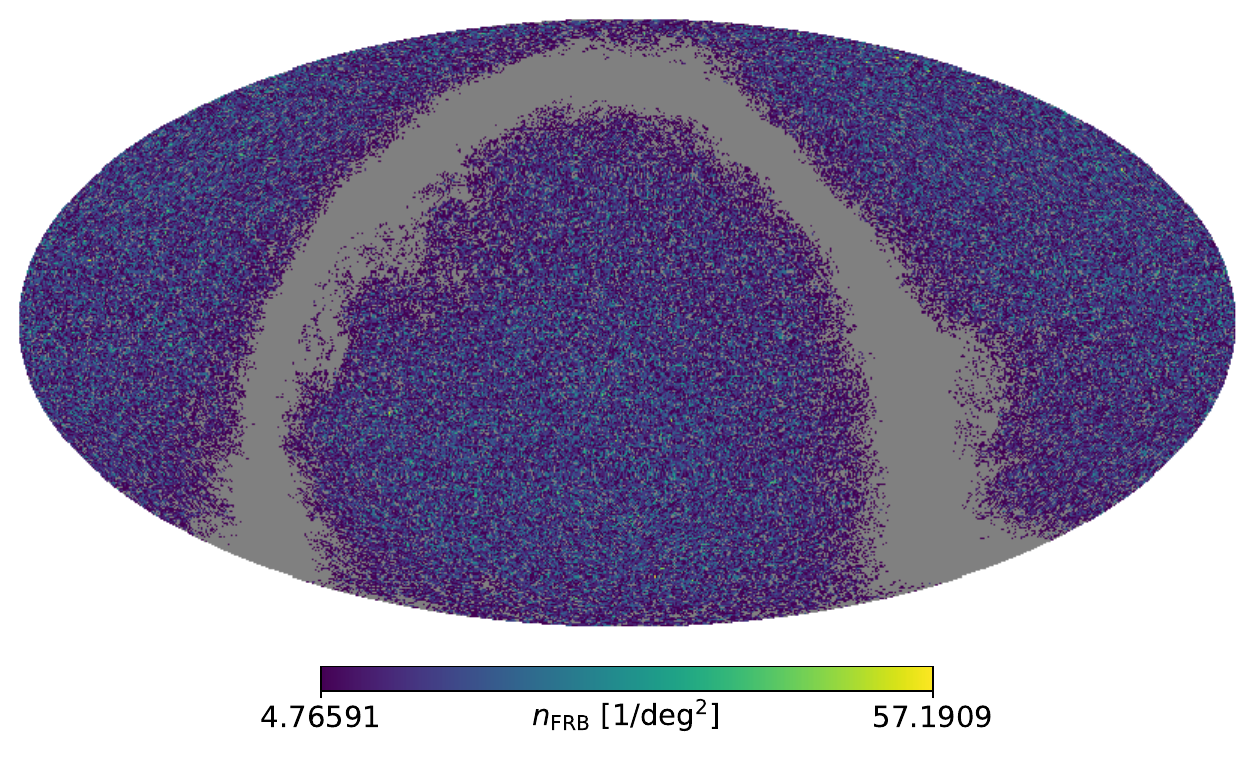}
        \includegraphics[width=0.44\textwidth]{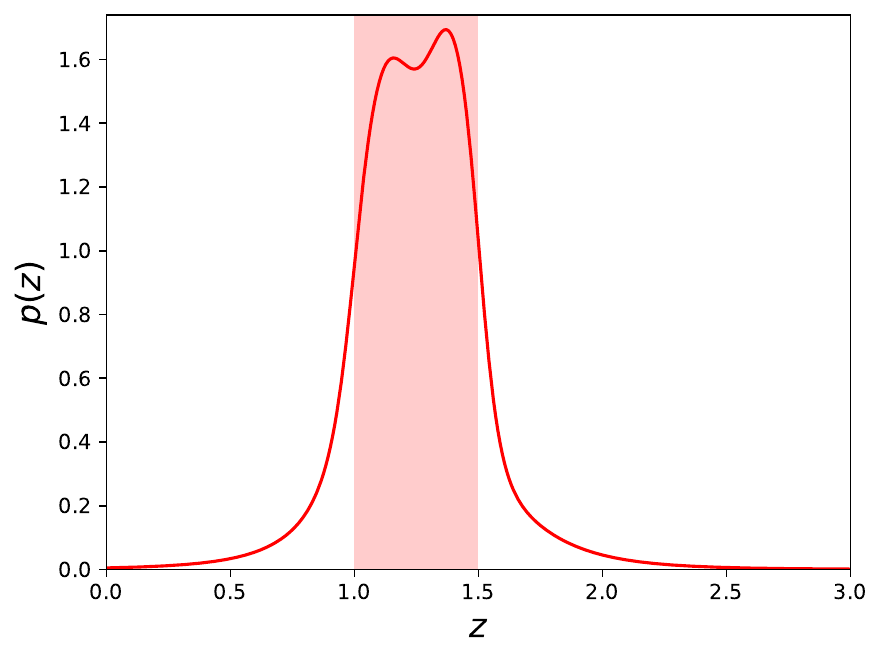}
        \caption{{\sl Left:} spatial distribution of the \emph{Quaia} quasar catalog used in this exercise, color-coded by its number density on the sphere. {\sl Right:} distribution of spectro-photometric redshift estimates for the same sample, with the vertical band indicating the redshift range of selected sources used in this work.}
        \label{fig:pm_data}
    \end{figure}
    Although most quantities of cosmological interest are either scalars or spin-2 objects (e.g. cosmic shear, CMB polarization), there are examples of particularly useful spin-1 fields, especially in terms of catalog-based observations. Examples of these are lensing displacements, galaxy angular momenta, and transverse peculiar velocities \cite{2110.01620,2305.15893}. Astrometric surveys, such as the \gaia satellite, are able to track the angular position of sources over time with high precision. Although the main use of such observations is reconstructing the dynamics of stars in the Milky Way, it has been argued that they may be useful for cosmology when applied to extra-galactic sources, placing constraints on the local rate of expansion \citep{1310.0500,1807.06658}, detecting the Solar System's proper motion dipole \citep{1811.05454}, and even constraining structure growth \citep{2305.15893}.

    As an example application, we generate simulated observations of quasar proper motions for a \gaia-like sample. In particular, we use the positions of quasars in the public \quaia quasar sample \cite{2306.17749}. We select 342{,}971 quasars with spectro-photometric redshifts in the range $1<z_{\rm sp}<1.5$. The spatial and redshift distribution of this sample is shown in Fig. \ref{fig:pm_data}. Next, we generate the theoretical prediction for the angular power spectrum of the expected cosmological proper motion signal. Following the formalism in \cite{2305.15893}, the projected proper motion map $\dot{\bf A}$ is a spin-1 field given by
    \begin{equation}
     \dot{\bf A}(\nv)=\int \frac{dz}{(1+z)\chi}p(z)\,(1-\nv^T\nv)\cdot{\bf v}(\chi\nv),
    \end{equation}
    where $p(z)$ is the redshift distribution of the sample, ${\bf v}$ is the peculiar velocity, projected on the plane transverse to the line of sight via the projector $1-\nv^T\nv$. At linear order in perturbation theory, $\dot{\bf A}$ is a pure $E$-mode field, since it can be written as the covariant derivative of a projected velocity potential \citep{2305.15893}. Thus, in what follows, we study only the power spectrum of its $E$-mode component.

    We compute a theoretical prediction for the signal power spectrum using the redshift distribution of the sample and assuming standard $\Lambda$CDM cosmological parameters. We then generate Gaussian realizations of this field at resolution $N_{\rm side}=1024$, and sample them at the positions of the \quaia quasars. Since the peculiar motion signal is most relevant at large scales, and given the relative sparsity of the sample, we will only attempt to reconstruct the power spectrum up to scales $\ell=360$, and thus this resolution is sufficient. We generate 100 such simulated catalogs, estimate their power spectrum using the catalog-based method presented here and the standard pixel-based approach, using maps of resolution $N_{\rm side}=128$ in the latter case.
    \begin{figure}
        \centering
        \includegraphics[width=0.7\textwidth]{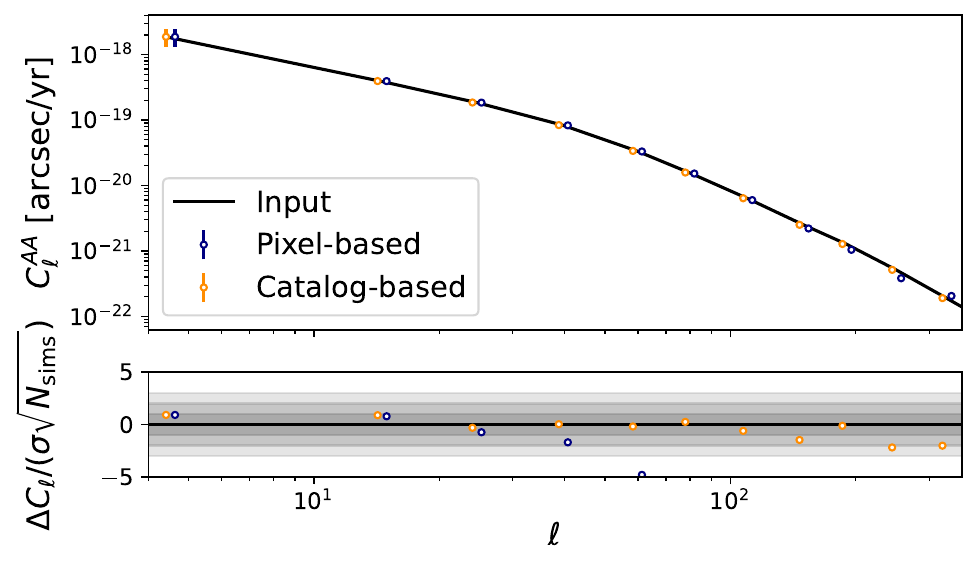}
        \caption{Angular power spectrum from 100 simulated \emph{Quaia}-like quasar proper motion catalogs, comparing the catalog-based (orange error bars) and the pixel-based approach (blue error bars, including a horizontal offset for clarity). The theoretical input is shown in black, and the relative difference with respect to it as a fraction of the statistical error on the mean of all 100 simulations is shown in the lower panel.}
        \label{fig:pm_cls}
    \end{figure}

    The result of this exercise is shown in Fig. \ref{fig:pm_cls}. The upper panel shows the $E$-mode power spectrum bandpowers measured and averaged over 100 simulations, as well as the associated theoretical prediction, given by the input power spectrum convolved with the bandpower window functions. The lower panel shows the relative residuals between simulations and theory prediction in both cases. We find that, while the catalog-based approach is able to obtain an unbiased estimate of the power spectrum at all scales, the pixel-based method under-predicts it severely on small scales ($\ell\gtrsim70$).

  \subsection{Spin-0 example: FRB-based dispersion measure maps} \label{ssec:ex.frb}
    As an example of a scalar (spin-0) quantity sampled at discrete positions, we study the case of maps of the projected electron density distribution constructed from the dispersion measure of fast radio bursts (FRBs). FRBs are powerful millisecond-long radio pulses emitted by (most likely) extra-galactic sources of unknown nature. As they propagate through the interstellar and intergalactic plasma, the group velocity of these pulses changes with frequency. This leads to a time delay between pulses at different frequencies $\nu$ that scales as $\Delta t=t_0\,{\rm DM}/\nu^2$, where $t_0\equiv 4.15\times10^{-3}\,\,{\rm s}\,{\rm GHz}^2\,{\rm cm}^3\,{\rm pc}^{-1}$, and the dispersion measure ${\rm DM}$ is
    \begin{equation}
      {\rm DM}(\nv)\equiv\int_0^{\chi_s}\frac{d\chi}{a}n_e(\chi\nv),
    \end{equation}
    with $\chi_s$ the comoving radial distance to the source, $a$ the scale factor, and $n_e({\bf x})$ the comoving electron density. As in the case of cosmic shear and peculiar transverse velocities, one can thus make use of FRB catalogs to construct maps of the projected electron density. For a catalog with a redshift distribution $p(z)$, this is given by \cite{astro-ph/0309200, astro-ph/0309364, arXiv:1401.0059}
    \begin{equation}\label{eq:DMfield}
      {\rm DM}(\nv)=\int \frac{dz}{H(z)}(1+z)\,P(>z)\,n_e(\chi(z)\nv)\, ,\hspace{12pt} P(>z)\equiv\int_z^\infty dz'\,p(z').
    \end{equation}
    where $P(>z)$ is the cumulative redshift distribution.
    \begin{figure}
        \centering
        \includegraphics[width=0.55\textwidth]{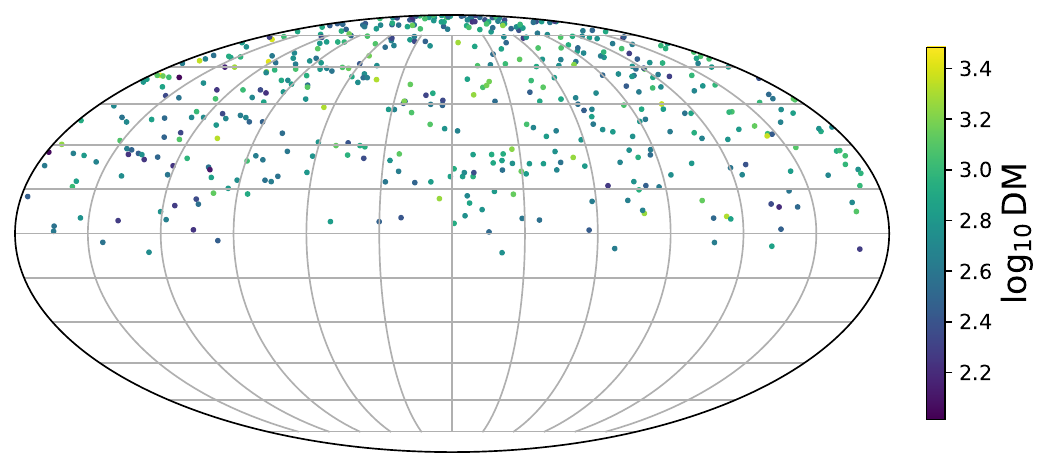}
        \includegraphics[width=0.43\textwidth]{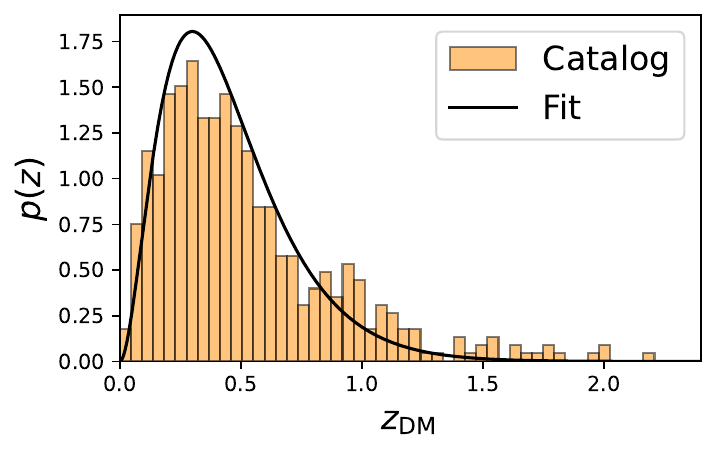}
        \caption{{\sl Left:} spatial distribution of the FRB catalog used in this exercise, with the color of each source displaying its measured DM. {\sl Right:} distribution of DM-based redshift estimates for the same sample, and fitting function used to generate the input angular power spectrum used in this study.}
        \label{fig:dm_data}
    \end{figure}

    Current FRB catalogs are significantly sparser than the transverse velocity and cosmic shear examples from previous sections, with existing FRB samples containing a few hundred sources (although this number will grow significantly in the future \cite{2107.10113}). Another distinctive feature of FRB catalogs is the fact that DM measurements are highly accurate. Typically, the DM can be measured with a precision of $\sim0.1\,{\rm cm}^{-3}{\rm pc}$, while extragalactic DM values are orders of magnitude higher (e.g. ${\rm DM}_{z\sim0.8}\sim1000\,{\rm cm}^{-3}\,{\rm pc}$), and the contributions from the Milky Way are also comparatively small (${\rm DM}_{\rm MW}\sim\,50\,{\rm cm}^{-3}{\rm pc}$). Typical host contributions are likely of the order ${\rm DM}_{\rm host}\sim100\,{\rm cm}^{-3}\,{\rm pc}$ (although the distribution of host DM may have significant tails \cite{2403.08611}). Thus, in this case, the per-source measurements of the field are signal-dominated, unlike the previous examples.

    To showcase the ability of the method presented here to recover accurate power spectrum measurements in this regime, we use the latest public catalog of FRBs made available by the CHIME collaboration \cite{2106.04352}. After removing duplicates, the catalog contains 492 unique sources, with DM values in the range ${\rm DM}\lesssim2000$. The left panel of Fig. \ref{fig:dm_data} shows the spatial distribution of this sample and their measured dispersion measures. We assign a nominal redshift to each source from their measured DM using the background relation between DM and redshift:
    \begin{equation}
      \overline{\rm DM}(z)-\overline{\rm DM}_0=\int_0^z\frac{dz'}{H(z')}\,(1+z')\,\bar{n}_e,\hspace{12pt}\bar{n}_e=\frac{3H_0^2\Omega_b}{8\pi G\,m_p}\frac{x_e(1+x_H)}{2},
    \end{equation}
    where $m_p$ is the proton mass, $H_0$ is the expansion rate today, $\Omega_b$ is the fractional density of baryons (which we set to $\Omega_b=0.05$), $x_e$ is the ionization fraction (which we set to $x_e=1$), and $x_H$ is the hydrogen mass fraction (which we set to $x_H=0.75$). We fix the offset $\overline{\rm DM}_0=100\,{\rm cm}^{-3}\,{\rm pc}$ to account for the typical host / Milky Way contribution (although our results do not depend on this choice). The corresponding redshift distribution of the sample is shown in the right panel of Fig. \ref{fig:dm_data}, and can be approximated as $p(z)\propto z^2\exp(-z/0.15)$ (solid line in the same figure).
    \begin{figure}
        \centering
        \includegraphics[width=0.49\textwidth]{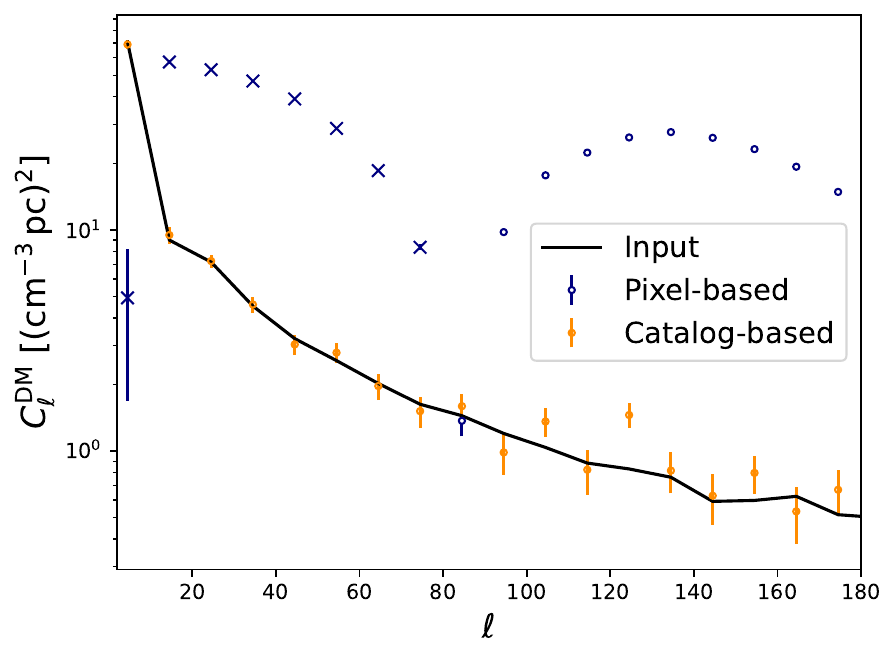}
        \includegraphics[width=0.49\textwidth]{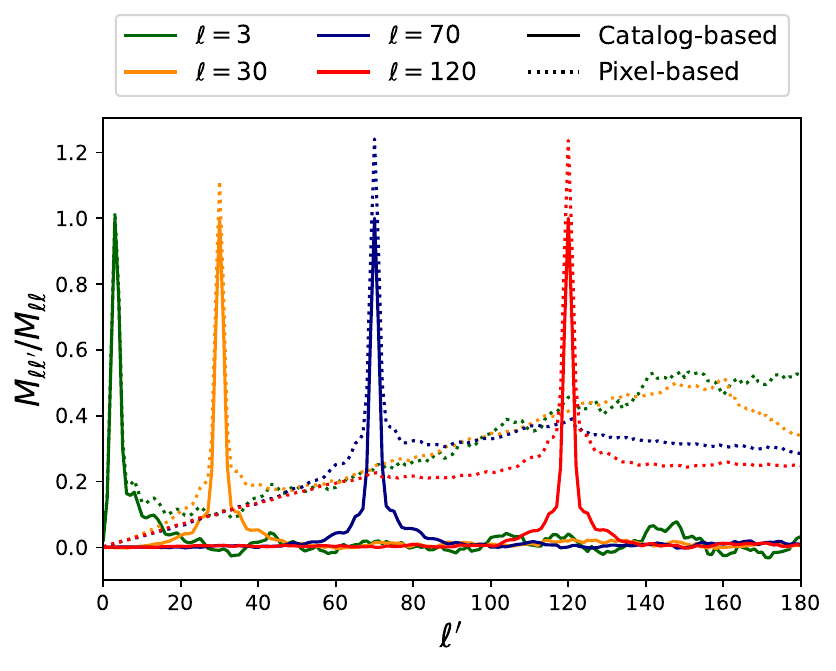}
        \caption{Results of the spin-0 FRB-based analysis. {\sl Left:} power spectrum, estimated using the catalog-based approach, averaged over 1000 Gaussian realizations (orange points), with error bars corresponding to the uncertainty on the mean $C_\ell$, and theoretical expectation (solid black line), corresponding to the input power spectrum convolved with the bandpower window functions. In blue, we show the result of the standard pixel-based approach, which fails catastrophically. {\sl Right:} rows of the mode-coupling matrix $M_{\ell \ell'}$ for different values of $\ell$, as a function of $\ell'$. Results are shown for the catalog-based and pixel-based approaches (solid and dotted lines, respectively).}
        \label{fig:dm_cls}
    \end{figure}

    We use this redshift distribution to generate a theoretical prediction for the power spectrum of the DM field in Eq. \ref{eq:DMfield}, assuming a linear bias for the electron overdensity of $b_e=0.8$ \cite{1810.13423}. We generate a suite of 1000 simulated Gaussian realizations of ${\rm DM}(\nv)$ at a \hp resolution of $N_{\rm side}=512$, which we then sample at the positions of the CHIME catalog sources. Given the sparsity of the sample, we only attempt to reconstruct the power spectrum up to scales $\ell=180$, and hence this reduced resolution is sufficient. The left panel of Fig. \ref{fig:dm_cls} shows, in orange error bars, the reconstructed power spectrum averaged over all simulations, together with its theoretical prediction (i.e., the bandpower-convolved power spectrum used to generate the simulated maps) in solid black lines. In blue error bars, we show also the result of using the standard, pixelized version of the estimator, using a resolution $N_{\rm side}=64$. The numerical instability of the mode-coupling matrix, and aliasing due to limited pixel resolution, causes the standard estimator to fail catastrophically in this case.
    
    The right panel of the same figure shows rows of the normalized mode-coupling matrix $M_{\ell\ell'}$ for four different values of $\ell$. Results are shown for the catalog-based and pixel-based approaches as solid and dotted lines, respectively. In the catalog-based case, $M_{\ell\ell'}$ displays significant coupling between multipoles separated by up to $\Delta\ell\sim20$, with a more non-trivial structure at larger multipole separations. In turn, the pixel-based approach, which does not correct for the effect of mask shot noise, shows significant correlation extending across the whole range of multipoles, growing like $(2\ell'+1)$ (see Eq. \ref{eq:mcm_N}). The mode-coupling matrix in this case is thus significantly more numerically unstable. Together with the impact of aliasing (see Appendix \ref{app:bias}), this leads to the failure observed in the right panel.

  \subsection{Clustering example}\label{ssec:ex.clust}
    We end this section showcasing the applicability of the discrete pseudo-$C_\ell$ approach to the case of source clustering, in which the density distribution of sources is itself the field under study. The formalism for this case was first presented in \cite{2312.12285}, and we described it within the general framework presented here in Section \ref{ssec:meth.clust}. To validate our implementation of the formalism in \nmt, we generate fast simulations of a galaxy clustering sample as follows:
    \begin{itemize}
      \item We calculate an initial angular power spectrum corresponding to the projected ovedensity of sources with a linear galaxy bias $b=1$ and a normal redshift distribution centered at $z=0.6$ with a standard deviation $\sigma_z=0.1$.
      \item We create a completeness map $c(\nv)$ to simulate the observational survey inhomogeneities that the random catalogs described in Section \ref{ssec:meth.clust} aim to trace. To do this, we draw 200 random unit vectors and select a spherical cap with a 10$^\circ$ radius centered at each of them. For each cap, we then draw a random number between 0.97 and 1 to represent the sky completeness in that patch (i.e., up to 3\% incompleteness). We multiply the resulting map by a sky mask that removes regions of high dust extinction and star contamination, constructed as described in \cite{1909.09102}.
      \item We generate a Gaussian overdensity field, $\delta_G(\nv)$ from this power spectrum, and perform a lognormal transformation to turn it into a positive-definite angular density field:
      \begin{equation}
        n^{\rm true}_g(\nv)=\bar{n}_g\,\exp(\delta_G(\nv)-\sigma_G^2/2),
      \end{equation}
      where $\sigma_G$ is the standard deviation of $\delta_G$, and $\bar{n}_g$ is the desired average number of sources per steradian\footnote{The $-\sigma_G^2/2$ counterterm ensures that the mean of the resulting field is indeed $\bar{n}_g$, assuming $\delta_G$ is Gaussian.}. The observed mean source density is then constructed by multiplying this field by the completeness map described above: $n^{\rm obs}_g(\nv)=c(\nv)\,n^{\rm true}_g(\nv)$.
      \item We Poisson-sample this field, assigning each source a random position within its pixel, to generate the ``data'' catalog. The ``random'' catalog is generated by drawing random positions on the sphere, and discarding them with a probability given by the local value of the completeness map. The random catalog thus constructed has 50 times more sources than the data.
    \end{itemize}
    \begin{figure}
    \centering
     \includegraphics[width=0.49\textwidth]{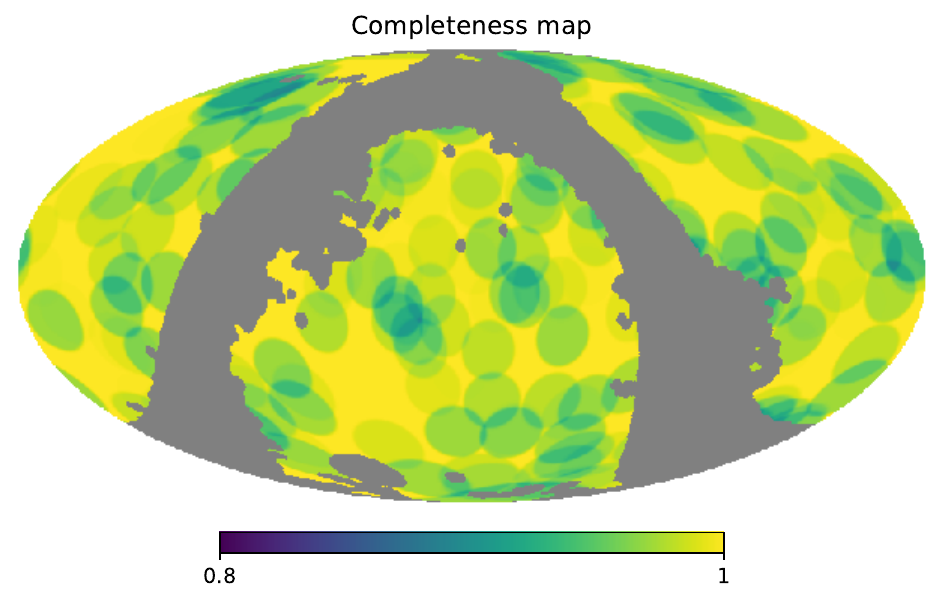}
     \includegraphics[width=0.49\textwidth]{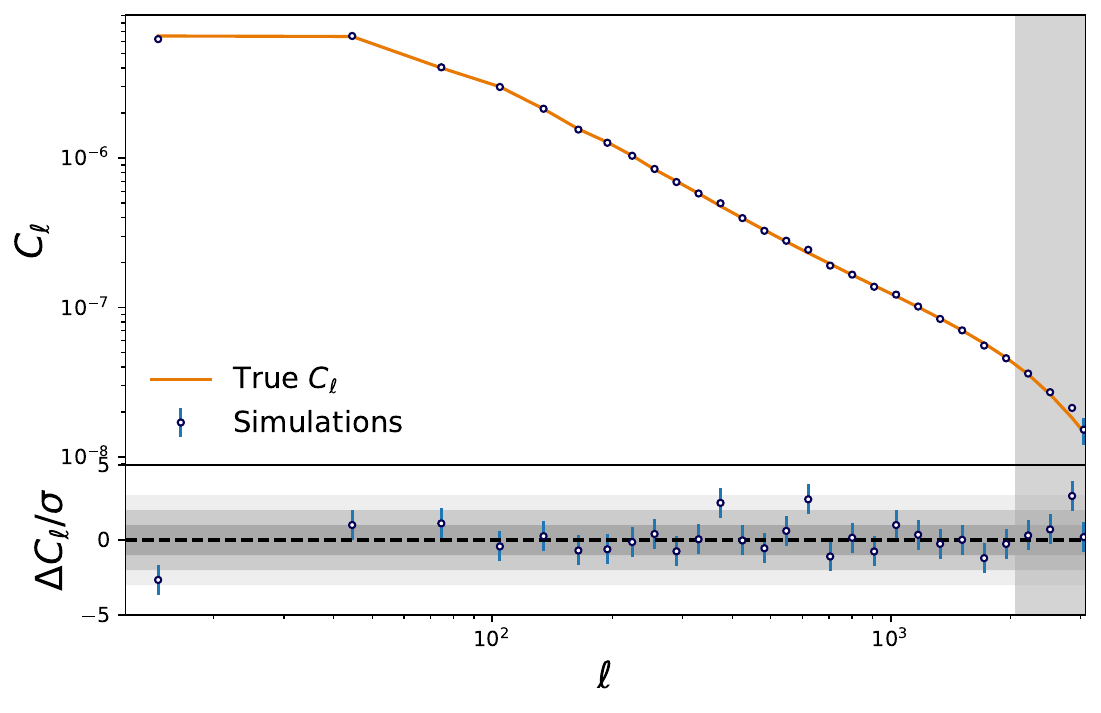}
    \caption{{\sl Left:} completeness map used for the clustering study described in Section \ref{ssec:ex.clust}. Fully masked pixels are shown in gray. {\sl Right:} power spectrum averaged over 100 simulations (dark blue circles with error bars), true underlying spectrum (orange line), and residual relative to the statistical uncertainties in the mean (bottom panel).} \label{fig:cls_clust}
    \end{figure}

    Keeping the completeness map and associated random catalog fixed, we generate 100 data simulations at $N_{\rm side}=1024$ with a number density $\bar{n}_g=0.01\,{\rm arcmin}^{-2}$, and estimate the power spectrum of each simulation using the procedure described in Section \ref{ssec:meth.clust}. The use of a lognormal transformation to generate the galaxy density field prevents us from using the initial angular power spectrum as a theoretical prediction to compare against. Although the effect of a lognormal transformation can be propagated to the power spectrum analytically (see \cite{1991MNRAS.248....1C}) we resort to a more direct approach that avoids potential numerical instabilities of Hankel transforms: for each simulation, we calculate the power spectrum of the lognormal density field ($\delta_{\rm LN}=\exp(\delta_G-\sigma_G^2/2)-1$) on the full sky (i.e., in the absence of completeness, masking, and discrete-source effects). The ``true'' power spectrum against which we compare the result of our simulations is then given by the average of this power spectrum over simulations corrected for the impact of the \hp pixel window function used in these simulations.

    The left panel of Fig. \ref{fig:cls_clust} shows the completeness map used in this exercise. The right panel of the same figure shows the mean power spectrum recovered from our simulation (blue points with error bars), the true $C_\ell$ (orange line), and the relative residual between them (bottom subpanel). The catalog method implemented here is able to recover the true underlying power spectrum at very high accuracy (the error bars shown represent the statistical uncertainties in the mean, corresponding to $1/10$ of the error of a single realization).

\section{Conclusion} \label{sec:con}
  In this work, we present a simple extension to the standard pseudo-$C_\ell$ estimator to tackle the problem of fields sampled at the positions of a discrete and finite catalog of sources. By treating the problem directly at the catalog level, without resorting to an intermediate pixelization of the sources onto a regular grid, we demonstrate that the method is able to tackle three prevalent sources of systematic bias in the standard pixel-based approach: resolution and aliasing effects due to the finite pixel area, shot noise in the field estimate due to the discrete nature of the field tracers, and numerically unstable mode-coupling matrices due to shot noise in the pixelized mask constructed from the sources. We interpret the catalog data as representing a masked field, with a mask given by a set of delta functions at the position of each source, and take advantage of new methods developed for the calculation of spherical harmonic transforms on irregular grids. We show that a simple correction of the shot-noise contributions to the power spectra of the masked field and of the mask itself, using fast and model-independent analytical expressions, is able to significantly improve the numerical stability of the mode-coupling matrix, while simultaneously rendering the estimator immune to any uncorrelated noise bias, irrespective of the size of the signal contribution. The method is a generalization of the technique presented in \cite{2312.12285} in the context of galaxy clustering. The most evident use of this method is in the analysis of cosmic shear data, although it is broadly applicable to any datasets described by observed properties of discrete sources.
  
  We validate our implementation of this method on a set 500 simulated KiDS-1000-like cosmic shear catalogs, with sources sampled at their real sky positions. We show that the resulting estimator is unbiased even for unrealistically small samples (100 and 10{,}000 times less dense than the real KiDS-1000 sample), where mode-coupling instabilities due to shot noise render standard pixel-based methods unusable. Even at realistic sample sizes, including the larger densities achieved by deeper samples, we show that pixel-based methods encounter difficulties to accurately characterize the effects of pixelization on the small-scale power spectrum. While these can be circumvented through the use of higher pixel resolutions, this is at a significantly higher cost in computational time and complexity, which the catalog-based method presented here avoids altogether.

  To demonstrate the practical use of our estimator on other suitable astrophysical datasets, with varying sample sizes and signal-to-noise ratios, we apply it in four additional distinct scenarios. First, we demonstrate its feasibility in the analysis of significantly larger catalogs, reproducing the analysis of the 3-year DES cosmic shear sample, comparing the results to those presented in \cite{2403.13794}. We then study the case of a spin-1 field: the peculiar transverse motion of extragalactic objects as traced by an astrometric survey, using the \quaia quasar sample. Besides the different spin, this type of dataset features a significantly lower source density and signal-to-noise ratio than cosmic shear. We also consider measurements of the spin-0 projected electron density in the form of Fast Radio Burst catalogs. These correspond to samples with extremely low number densities ($10^2$-$10^3$ objects) but a very high signal-to-noise ratio. Finally, we showcase the applicability of our formalism to the analysis of galaxy clustering, in which the density of catalog sources is itself the field of interest.  Our results from all four examples confirm our previous findings that our estimator is unbiased and can overcome the numerical challenge of source sparseness when characterising mode coupling.

  We expect the method presented to be useful in the analysis of upcoming large-scale structure datasets, particularly in the context of cosmic shear, as well as novel probes such as FRBs, radial and transverse peculiar motion surveys, weak lensing of gravitational waves, and any other catalog-based observations. Our implementation is made available with the latest release of the public power spectrum package \nmt\footnote{\url{https://github.com/LSSTDESC/NaMaster/}}. 

\section*{Acknowledgments}
  When the work presented here was at an advanced stage, we learnt of similar work being developed by Tessore et al. 2024 \cite{TessoreInPrep}. We thank the authors of this work, particularly Nicolas Tessore and Benjamin Joachimi, for sharing their draft paper and for useful discussions. We also thank Anton Baleato, Elyas Farah, Robert Reischke, and Martin White, for useful comments and feedback. KW acknowledges support from the Science and Technology Facilities Council (STFC) under grant ST/X006344/1. DA acknowledges support from the Beecroft trust. We made extensive use of computational resources at the University of Oxford Department of Physics, funded by the John Fell Oxford University Press Research Fund.

\appendix
\section{Generalities of spherical harmonic functions}\label{app:shts}
  We use the following definition for the generalized spin-$s$ spherical harmonic functions\footnote{The $-1$ sign preceding these definitions is a mostly inconvenient convention that is dropped for the case of spin-0 fields, so that $_0Y^{1E}_{\ell m}=\,_0Y^{2B}_{\ell m}=Y_{\ell m}$, and $_0Y^{2E}_{\ell m}=\,_0Y^{1B}_{\ell m}=0$.}:
  \begin{align}
    _sY^{1E}_{\ell m}=-\frac{1}{2}[\,_sY_{\ell m}+(-1)^s\,_{-s}Y_{\ell m}],\hspace{12pt}
    _sY^{1B}_{\ell m}=-\frac{i}{2}[\,_sY_{\ell m}-(-1)^s\,_{-s}Y_{\ell m}],\\
    _sY^{2E}_{\ell m}=\frac{i}{2}[\,_sY_{\ell m}+(-1)^s\,_{-s}Y_{\ell m}],\hspace{12pt}
    _sY^{1B}_{\ell m}=-\frac{1}{2}[\,_sY_{\ell m}-(-1)^s\,_{-s}Y_{\ell m}],
  \end{align}
  where $_sY_{\ell m}(\nv)$ are the spin-weighted spherical harmonic functions of spin $s$.
  These functions satisfy the following properties:
  \begin{align}
    &\int d\nv\,_sY_{\ell m}^{a\alpha*}(\nv)_sY_{\ell' m'}^{a\beta}(\nv)=\delta_{\alpha\beta}\delta_{\ell\ell'}\delta_{mm'},\\\nonumber
    &\int d\nv\,_sY_{\ell m}^{a\alpha*}(\nv)_sY_{\ell' m'}^{a\beta}(\nv)\,Y_{\ell''m''}(\nv)=(-1)^{s+m}\sqrt{\frac{(2\ell+1)(2\ell'+1)(2\ell''+1)}{4\pi}}\\
    &\hspace{175pt}\wtj{\ell}{\ell'}{\ell''}{-m}{m'}{m''}\wtj{\ell}{\ell'}{\ell''}{s}{-s}{0}\,d^{\alpha\beta}_{\ell+\ell'+\ell''},\\
    &d^{EE}_n=d^{BB}_n=[1+(-1)^n]/2,\hspace{12pt}d^{EB}_n=-d^{BE}_n=-i[1-(-1)^n]/2.\\\label{eq:sYlm_add}
    &\sum_m\,_sY_{\ell m}^{a\alpha*}(\nv)\,_{s'}Y_{\ell m}^{a\beta}(\nv)=\frac{2\ell+1}{4\pi}\delta_{\alpha\beta}\delta_{ss'}.
  \end{align}
  The last relation can be derived from the addition theorem for spin-weighted spherical harmonics \cite{siebert_thesis}. For scalar harmonics, it reads simply
  \begin{equation}\label{eq:Ylm_add}
    \sum_m Y^*_{\ell m}(\nv_1)Y_{\ell m}(\nv_2)=\frac{2\ell+1}{4\pi} P_\ell(\nv_1\cdot\nv_2) \, ,
  \end{equation}
  which, for $\nv_1 = \nv_2$, simply evaluates to $(2\ell+1)/(4\pi)$. Furthermore, the spherical harmonic functions are a complete basis on the sphere, so that:
  \begin{equation}\label{eq:Ylm_comp}
    \sum_{\ell m}Y^*_{\ell m}(\nv_1)Y_{\ell m}(\nv_2)=\delta^D(\nv_1,\nv_2).
  \end{equation}
  We will also employ the following properties of the Wigner $3j$ symbols:
  \begin{align}
    &\sum_{mm'}\wtj{\ell}{\ell'}{\ell''}{m}{m'}{m''}\wtj{\ell}{\ell'}{\ell'''}{m}{m'}{m'''}=\frac{\delta_{\ell''\ell'''}\delta_{m''m'''}}{2\ell''+1},\\\label{eq:w3j_orth2}
    &\sum_{\ell''}(2\ell''+1)\wtj{\ell}{\ell'}{\ell''}{s}{-s}{0}\wtj{\ell}{\ell'}{\ell''}{s'}{-s'}{0}=\delta_{ss'},\\\label{eq:w3j_flip}
    &\wtj{\ell}{\ell'}{\ell''}{-m}{-m'}{-m''}=(-1)^{\ell+\ell'+\ell''}\wtj{\ell}{\ell'}{\ell''}{m}{m'}{m''}.
  \end{align}

\section{Pixelization effects in catalog-based observables}\label{app:bias}
  The main text has emphasized the impact of the numerical instability of the mode-coupling matrix due to the presence of shot noise in the mask power spectrum for catalog-based observables that are first pixelized before estimating their angular power spectrum. This appendix explores the two other sources of potential bias that affect pixelized catalog-based data: the impact of pixel window functions, and aliasing.
  \begin{figure}
    \centering
    \includegraphics[width=0.99\textwidth]{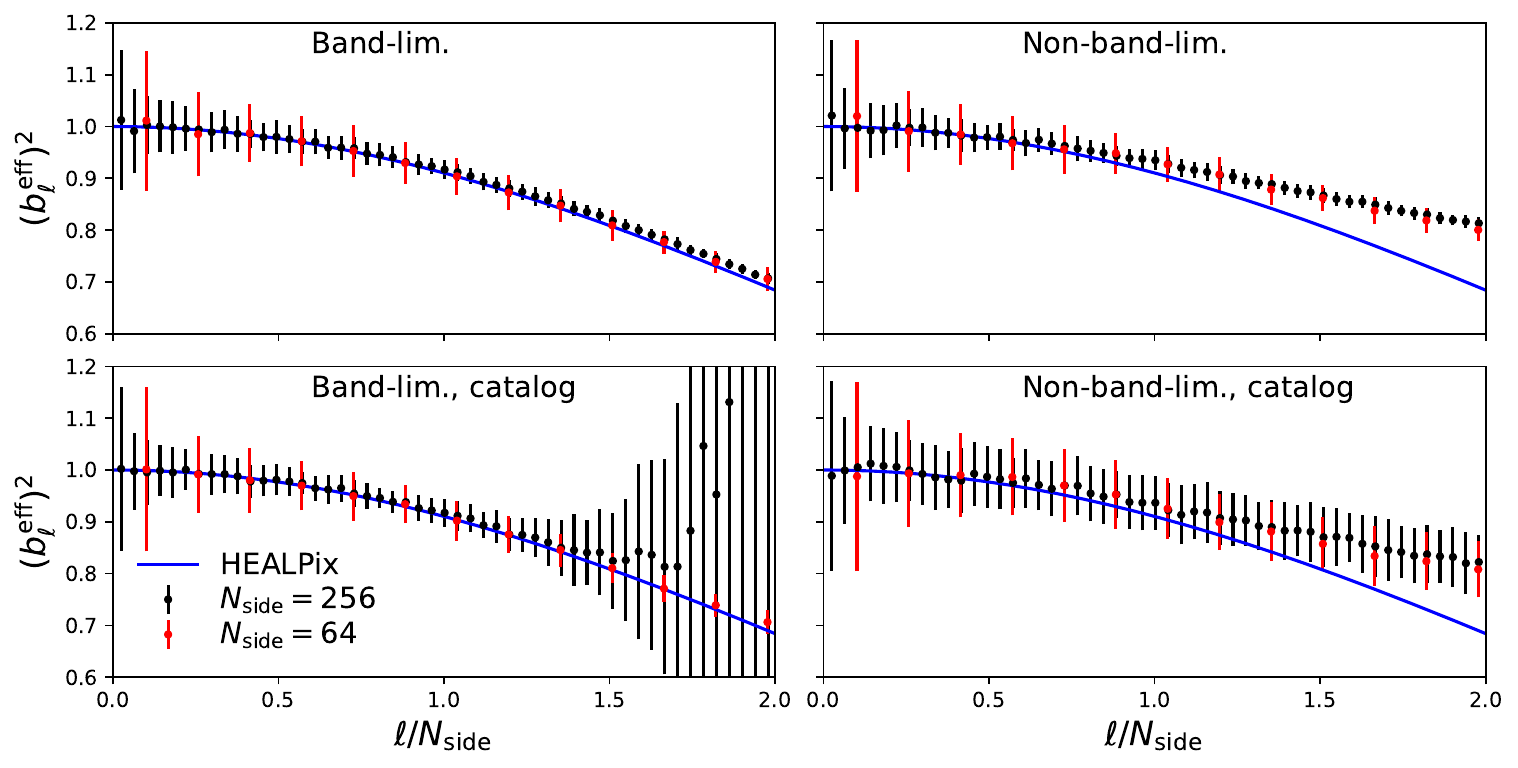}
    \caption{Effective pixel window function for band-limited simulations (left panels) and non-band-limited simulations (right panels). Results for medium- and low-resolution pixels ($N_{\rm side}=256$ and 64, respectively) are shown as black and red points with error bars, compared against the \hp window function (blue). The top panels show results for simulations in which the base maps at $N_{\rm side}=1024$ are averaged down to the lower resolutions. The bottom panels show results for catalogs constructed from the base maps and then pixelized at lower resolution. Besides the numerical instability of the mode-coupling matrix for catalog-based measurements, which is factored out in this exercise, the most relevant residual resolution effect relevant for map-based analyses of catalog datasets is aliasing.}
    \label{fig:aliasing}
  \end{figure}

  For this exercise we generate two types of simulated realizations of a  scalar Gaussian random field using a base pixelization of $N_{\rm side}=1024$:
  \begin{itemize}
      \item {\bf Band-limited realizations} with a power spectrum of the form
      \begin{equation}\label{eq:clin_bl}
        C_\ell=\frac{\exp[-(\ell\Theta)^2]}{\ell+50}.
      \end{equation}
      The exponential cutoff, with $\Theta=0.3^\circ$ sharply reduces any structure on scales $\ell\gtrsim200$.
      \item {\bf Non-band-limited realizations} with a power spectrum of the form
      \begin{equation}\label{eq:clin_nbl}
        C_\ell=\frac{1}{\ell+50},
      \end{equation}
      which is rather flat, decaying significantly more slowly at high $\ell$ than the previous case.
  \end{itemize}

  First, to isolate the effects of pixel averaging and aliasing from the practicalities of fields sampled at arbitrary discrete positions, we construct maps at lower resolution, $N_{\rm side}=256$ and $N_{\rm side}=64$ by simply averaging the $N_{\rm side}=1024$ maps onto larger pixels. We calculate the angular power spectrum of each of these maps, and estimate the effective pixel window function of the measured power spectrum as:
  \begin{equation}
    b^{\rm eff}_\ell\equiv \sqrt{\frac{\hat{C}_\ell}{C_\ell^{\rm in}}},
  \end{equation}
  where $\hat{C}_\ell$ is the estimated power spectrum, and $C_\ell^{\rm in}$ is the input power spectrum (given by Eq. \ref{eq:clin_bl} or Eq. \ref{eq:clin_nbl}). We repeat this for 100 simulations and compute the mean and scatter of the effective pixel window function.

  The top left and top right panels of Fig. \ref{fig:aliasing} shows the result of this exercise for the band-limited and non-band-limited realizations, respectively. We find that, for the band-limited case, the impact of pixel averaging is captured by the \hp pixel window function with relatively good accuracy\footnote{The slight difference for the $N_{\rm side}=256$ maps is due to the fact that averaging over only $4\times4$ sub-pixels is less representative of a true area averaging than the $16\times16$ pixels for $N_{\rm side}=64$.}. In the case of non-band-limited simulations, however, we see very significant differences in the effective pixel window function. These differences are caused by aliasing of the small-scale power in the base maps onto the larger pixels via averaging. Interestingly, the resulting effective pixel window function has a similar shape both in the $N_{\rm side}=256$ and the $N_{\rm side}=64$ case.

  We now repeat the exercise with a key modification: we sample the base maps at the discrete positions of $10^6$ sources drawn at random on the sphere, and we then construct maps of the resulting field using the standard pixel-based approach at $N_{\rm side}=256$ and $N_{\rm side}=64$ (corresponding to $\sim30$ and $\sim1.3$ sources per pixel on average, respectively). When calculating the power spectra of theses maps we use the density of said sources as their corresponding masks. Finally, to factor out the effects caused by shot noise in the mask, we subtract the analytical shot noise contribution to the mask and field pseudo-$C_\ell$ ($\tilde{N}^w$ and $\tilde{N}^f$), mimicking the procedure described in the main text to avoid these effects. The bottom left and right panels of Fig. \ref{fig:aliasing} show the results of this exercise for the band-limited and non-band-limited maps, respectively. The conclusions are analogous to those found in the previous exercise: regardless of the average number of sources per pixel, the most relevant effect is aliasing in the case of non-band-limited data. The effects of pixelization on band-limited data are well captured by the \hp pixel window function.

\bibliography{references}{}

\end{document}